\documentclass[12pt,a4paper,english,draftcls, onecolumn]{IEEEtran}
\usepackage[T1]{fontenc}
\usepackage[latin9]{inputenc}
\usepackage{color}
\usepackage{babel}
\usepackage{verbatim}
\usepackage{float}
\usepackage{amsthm}
\usepackage{amsmath}
\usepackage{amssymb}
\usepackage{graphicx}
\usepackage{setspace}
\usepackage{esint}
\usepackage{epstopdf}

\makeatletter

\pdfpageheight\paperheight
\pdfpagewidth\paperwidth

\providecommand{\tabularnewline}{\\}

\theoremstyle{plain}
\newtheorem{thm}{\protect\theoremname}
\theoremstyle{definition}
\newtheorem{defn}[thm]{\protect\definitionname}
\theoremstyle{plain}
\newtheorem{cor}[thm]{\protect\corollaryname}

\usepackage{subfigure}
\usepackage{epstopdf}
\usepackage{cite}
\usepackage{citesort}
\usepackage{balance}

\makeatother

\providecommand{\corollaryname}{Corollary}
\providecommand{\definitionname}{Definition}
\providecommand{\theoremname}{Theorem}

\begin{document}

\title{Microscopic Analysis of the Uplink Interference in FDMA Small Cell
Networks
}

\author{\noindent {\normalsize{}Ming Ding, }\textit{\normalsize{}Member,
IEEE}{\normalsize{}, David L$\acute{\textrm{o}}$pez-P$\acute{\textrm{e}}$rez,
}\textit{\normalsize{}Member, IEEE}{\normalsize{}, Guoqiang Mao, }\textit{\normalsize{}Senior
Member, IEEE}{\normalsize{}, Zihuai Lin, }\textit{\normalsize{}Senior
Member, IEEE}%
\thanks{Ming Ding is with the National ICT Australia (NICTA) (e-mail: Ming.Ding@nicta.com.au).
David L$\acute{\textrm{o}}$pez-P$\acute{\textrm{e}}$rez is with
Bell Labs Alcatel-Lucent, Ireland (email: dr.david.lopez@ieee.org).
Guoqiang Mao is with the School of Computing and Communication, The
University of Technology Sydney, Australia and National ICT Australia
(NICTA) (e-mail: g.mao@ieee.org). Zihuai Lin is with the School of
Electrical and Information Engineering, The University of Sydney,
Australia (e-mail: zihuai.lin@sydney.edu.au).%
}
}

\maketitle
{}
\begin{abstract}
In this paper, we analytically derive an upper bound on the error
in approximating the uplink (UL) single-cell interference by a lognormal
distribution in frequency division multiple access (FDMA) small cell
networks (SCNs). Such an upper bound is measured by the Kolmogorov\textendash Smirnov
(KS) distance between the actual cumulative density function (CDF)
and the approximate CDF. The lognormal approximation is important
because it allows tractable network performance analysis%
. Our results are more general than the existing works in the sense
that we do not pose any requirement on (i) the shape and/or size of
cell coverage areas, (ii) the uniformity of user equipment (UE) distribution,
and (iii) the type of multi-path fading. Based on our results, we
propose a new framework to directly and analytically investigate a
complex network with practical deployment of multiple BSs placed at
irregular locations, using a power lognormal approximation of the
aggregate UL interference. The proposed network performance analysis
is particularly useful for the 5th generation (5G) systems with general
cell deployment and UE distribution. %
\footnote{1536-1276 © 2015 IEEE. Personal use is permitted, but republication/redistribution requires IEEE permission. Please find the final version in IEEE from the link: http://ieeexplore.ieee.org/document/7426421/. Digital Object Identifier: 10.1109/TWC.2016.2538261}

\end{abstract}

\section{Introduction}

Small cell networks (SCNs) have been identified as one of the key
enabling technologies in the 5th generation (5G) networks~\cite{Tutor_smallcell}.
In this context, new and more powerful network performance analysis
tools are being developed to gain a deep understanding of the performance
implications that SCNs bring about. These tools are significantly
different from traditional network performance analysis tools applicable
for studying just a few macrocells only. These network performance
analysis tools can be broadly classified into two large groups, i.e.,
macroscopic analysis and microscopic analysis~{[}2-10{]}.

The macroscopic analysis assumes that both user equipments (UEs) and
base stations (BSs) are randomly deployed in the network, often following
the homogeneous Poisson%
{} distribution, and usually try to derive the signal-to-interference-plus-noise
ratio (SINR) distribution of UEs and other performance metrics such
as the coverage probability and the area spectral efficiency~{[}2,3{]}.
The microscopic analysis is often conducted assuming that UEs are
randomly placed and BSs are deterministically deployed, i.e., the
BS positions are known~{[}4-10{]}. %

The microscopic analysis is important because it allows for a network-specific
study and optimization, e.g., optimizing the parameters of UL power
control~\cite{UL_interf_sim1} and performing per-cell loading balance
in a specific SCN~\cite{UL_interf_sim2}. In contrast, the macroscopic
analysis investigates network performance at a high level by averaging
out all the possible BS deployments~{[}2,3{]}. Generally speaking,
the microscopic analysis gives more targeted results for specific
networks than the macroscopic analysis, while the macroscopic analysis
gives a general picture of the network performance.

In this paper, we focus on the microscopic analysis. In particular,
we consider an uplink (UL) frequency division multiple access (FDMA)
SCN, which has been widely adopted in the 4th generation (4G) networks,
i.e., the UL single-carrier FDMA (SC-FDMA) system in the 3rd Generation
Partnership Project (3GPP) Long Term Evolution (LTE) networks~\cite{TS36.213}
and the UL orthogonal FDMA (OFDMA) system in the Worldwide Interoperability
for Microwave Access (WiMAX) networks~\cite{WiMAX_AI}. %
For the UL microscopic analysis, the existing works use
\begin{itemize}
\item Approach~1, which provides closed-form expressions but complicated
analytical results for a network with\emph{ a small number of interfering
cells and each cell has a regularly-shaped coverage area}, e.g., a
disk or a hexagon~\cite{UL_interf_2cells_ICC}. In~\cite{UL_interf_2cells_ICC},
the authors considered a single UL interfering cell with a disk-shaped
coverage area and presented closed-form expressions for the UL interference
considering both path loss and shadow fading.
\item Approach~2, which first analyzes the UL interference and then makes
an empirical assumption on the UL interference distribution, and on
that basis derives analytical results for a network with \emph{multiple
interfering cells, whose BSs are placed on a regularly-shaped lattice},
e.g., a hexagonal lattice~\cite{UL_interf_LN_conjecture_CDMA}-\hspace{-0.03cm}\cite{UL_interf_LN_conjecture_WCL}.
Specifically, in~\cite{UL_interf_LN_conjecture_CDMA} and~\cite{UL_interf_LN_conjecture_CDMA2},
the authors showed that the lognormal distribution better matches
the distribution of the uplink interference than the conventionally
assumed Gaussian distribution in a hexagonal cellular layout. %
In~\cite{UL_interf_LN_conjecture_WCL}, the authors assumed that
the uplink interference in hexagonal grid based OFMDA cellular networks
should follow a lognormal distribution. Such assumption was verified
via simulation.
\item Approach~3, which conducts system-level simulations to directly obtain
empirical results for a complex network with \emph{practical deployment
of multiple cells, whose BSs are placed at irregular locations}~\cite{Tutor_smallcell},~\cite{UL_interf_sim1},~\cite{UL_interf_sim2}.
In particular, the authors of \cite{Tutor_smallcell},~\cite{UL_interf_sim1},~\cite{UL_interf_sim2}
conducted system-level simulations to investigate the network performance
of SCNs in existing 4G networks and in future 5G networks.
\end{itemize}

Obviously, Approach~1 and Approach~3 lack generality and analytical
rigor, respectively. Regarding Approach~2, it has been a number of
years since an empirical conjecture was extensively used in performance
analysis, which stated that the UL inter-cell interference with \emph{disk-shaped}
coverage areas and uniform UE distributions could be well approximated
by a lognormal distribution in code division multiple access (CDMA)
SCNs~\cite{UL_interf_LN_conjecture_CDMA}, \cite{UL_interf_LN_conjecture_CDMA2}
and in FDMA SCNs~\cite{UL_interf_LN_conjecture_WCL}. This conjecture
is important since the lognormal approximation of interference distribution
allows tractable network performance analysis%
. However, up to now, it is still unclear \emph{how accurate} this
lognormal approximation is. %
In this paper, we aim to answer this fundamental question, and thus
making a significant contribution to constructing a formal tool for
the microscopic analysis of network performance. Note that in our
previous work~\cite{UL_interference_approx_GC}, we investigated
an upper bound on the error of this lognormal approximation under
the assumptions of uniform UE distribution and Rayleigh multi-path
fading. In this paper, we will largely extend our previous work by
presenting a new and tighter upper bound on the approximation error
and remove the requirement on the types of UE distribution and multi-path
fading.

In this paper we focus on the analysis of UL inter-cell interference.
Note that the interference analysis is important because it paves
way to the analyses of SINR, as well as other performance metrics
such as the coverage probability and the area spectral efficiency.
The contributions of this paper are as follows:
\begin{enumerate}
\item Our work %
analytically derives an upper bound on the error in approximating
the UL single-cell interference in FDMA SCNs by a lognormal distribution.
Such error is measured by the Kolmogorov\textendash Smirnov (KS) distance~\cite{KS-distance}
between the actual cumulative density function (CDF) and the approximate
CDF.
\item Unlike the existing works on the microscopic analysis, e.g.,~{[}4-11{]},
our work does not pose any requirement on (i) the shape and/or size
of cell coverage areas, (ii) the uniformity of UE distribution, and
(iii) the type of multi-path fading. Thus, our proposed framework
is more general and useful for network performance analysis.
\item Based on our work, a new approach can be established to fill an important
theoretical gap in the existing microscopic analysis of network performance,
which either assumes very simple BS deployments or relies on empirical
results. Such new approach allows us to directly investigate a complex
network with \emph{practical deployment of multiple BSs placed at
irregular locations}, while retaining mathematical rigor in the analysis.
In order to do that, we first verify the accuracy of the approximated
UL interference distribution for each small cell, then we approximate
the aggregate UL interference by a power lognormal distribution. Specifically,
the CDF of a power lognormal distribution is a power function of the
CDF of a lognormal distribution%
.
\end{enumerate}

The remainder of the paper is structured as follows. In Section~\ref{sec:Network-Model},
the network scenario and the system model are described. In Section~\ref{sec:Analysis-of-UL-Interf},
our approach to studying the UL inter-cell interference in FDMA SCNs
is presented, followed by the validation of our results via simulations
in Section~\ref{sec:Simulaiton-and-Discussion}. Finally, the conclusions
are drawn in Section~\ref{sec:Conclusion}.

\section{Network Scenario and System Model\label{sec:Network-Model}}

In this paper, we consider UL transmissions and assume that in one
frequency resource block (RB) and in a given time slot, only \emph{one}
UE is scheduled by each small cell BS to perform an UL transmission,
which is a reasonable assumption in line with the 4G networks, i.e.,
the UL SC-FDMA system in the LTE networks~\cite{TS36.213} and the
UL OFDMA system in the WiMAX networks~\cite{WiMAX_AI}. We assume
that each small cell has at least one associated UE because the small
cell BSs having no UE do not contribute to the uplink interference
analyzed in this paper, thereby can be ignored.

\begin{figure}[h]
\vspace{-0.5cm}

\noindent \begin{centering}
\includegraphics[width=7cm]{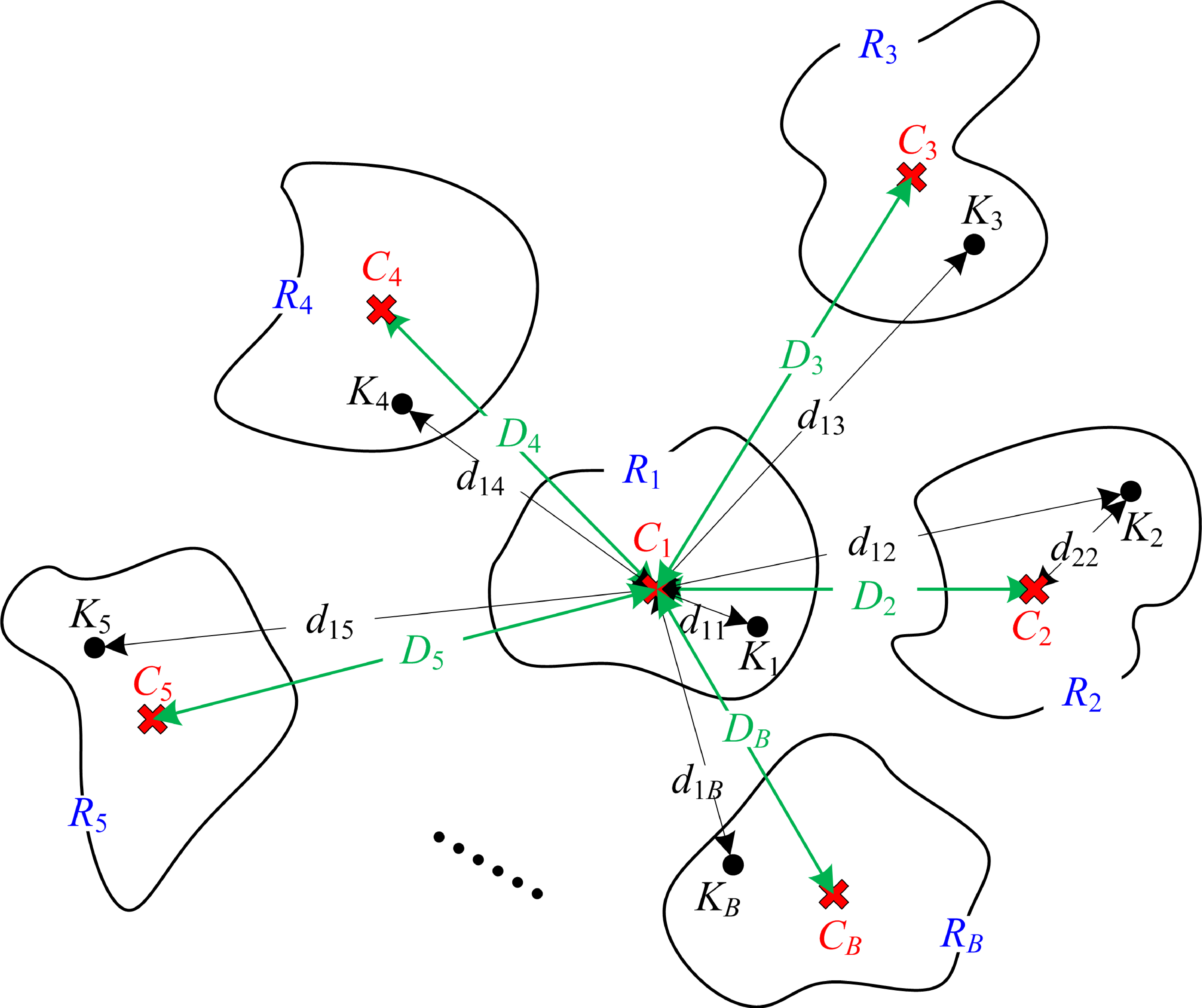} \renewcommand{\figurename}{Fig.}\protect\caption{\label{fig:sys_model}A schematic model of the considered SCN.}

\par\end{centering}

\vspace{-0.5cm}
\end{figure}

Regarding the network scenario, we consider a SCN with multiple small
cells operating on the same carrier frequency as shown in Fig.~\ref{fig:sys_model}.
In Fig.~\ref{fig:sys_model}, the SCN consists of $B$ small cells,
each of which is managed by a BS. The network includes one small cell
of interest denoted by $C_{1}$ and $B-1$ interfering small cells
denoted by $C_{b},b\in\left\{ 2,\ldots,B\right\} $. We focus on a
particular frequency RB, and denote by $K_{b}$ the active UE associated
with small cell $C_{b}$ using that frequency RB. Moreover, we denote
by $R_{b}$\emph{ the coverage area} of small cell $C_{b}$, in which
its associated UEs are randomly distributed. Note that the coverage
areas of adjacent small cells may overlap due to the arbitrary shape
and size of $\left\{ R_{b}\right\} ,b\in\left\{ 2,\ldots,B\right\} $.

\textit{The distance} (in km) from the BS of $C_{b}$ to the BS of
$C_{1}$, $b\in\left\{ 1,\ldots,B\right\} $, and the distance from
UE $K_{b}$ to the BS of $C_{m}$, $b,m\in\left\{ 1,\ldots,B\right\} $,
are denoted by $D_{b}$ and $d_{bm}$, respectively. In this paper,
we consider a deterministic deployment of BSs, and thus the set $\left\{ D_{b}\right\} $
is assumed to be known. However, UE $K_{b}$ is assumed to be randomly
distributed in $R_{b}$ with a distribution function $f_{Z_{b}}\left(z\right),z\in R_{b}$.
Hence, $d_{bm}$ is a random variable (RV), whose distribution cannot
be readily expressed in an analytical form due to the arbitrary shape
and size of $R_{b}$, and the arbitrary form of $f_{Z_{b}}\left(z\right)$.
Unlike the existing works, e.g.,~{[}2-8{]}, that only treat uniform
UE distributions, in this work we investigate a general probability
density function (PDF) of UE distribution denoted by $f_{Z_{b}}\left(z\right)$,
which satisfies $0<f_{Z_{b}}\left(z\right)<+\infty,z\in R_{b},$ and
its integral over $R_{b}$ equals to one, i.e., $\int_{R_{b}}f_{Z_{b}}\left(z\right)dz=1$.

In the following, we present the modeling of path loss, shadow fading,
UL transmission power, %
and multi-path fading, respectively.

Based on the definition of $d_{bm}$, \textit{the path loss} (in dB)
from UE $K_{b}$ to the BS of $C_{m}$ is modeled as

\begin{singlespace}
\noindent
\begin{equation}
L_{bm}={A}+{\alpha}{\log_{10}}{d_{bm}},\label{eq:PL_BS2UE}
\end{equation}

\end{singlespace}

\noindent where $A$ is the path loss at the reference distance of
$d_{bm}=1$ and $\alpha$ is the path loss exponent. In practice,
$A$ and $\alpha$ are constants obtainable from field tests~\cite{TR36.828}.
Note that $L_{bm}$ is a RV due to the randomness of $d_{bm}$.

\noindent %

\noindent %

\noindent %

\begin{singlespace}
\noindent %

\end{singlespace}

\textit{The shadow fading} (in dB) from UE $K_{b}$ to the BS of $C_{m}$
is denoted by $S_{bm}$%
, and it is usually modeled as a zero-mean Gaussian RV because the
linear-scale value of $S_{bm}$ is commonly assumed to follow a lognormal
distribution~\cite{TR36.828}. Hence, in this paper, we model $S_{bm}$
as an independently and identically distributed (i.i.d.) zero-mean
Gaussian RV with variance $\sigma_{\textrm{Shad}}^{2}$, denoted by
$S_{bm}\sim\mathcal{N}\left(0,\sigma_{\textrm{Shad}}^{2}\right)$%
{} .

\textit{The UL transmission power} (in dBm) of UE $K_{b}$ is denoted
by $P_{b}$. In practice, $P_{b}$ is usually subject to a semi-static
power control (PC) mechanism%
\footnote{\noindent Note that in practice $P_{b}$ is also constrained by the
maximum value of the UL power, denoted by $P_{\textrm{max}}$ at the
UE. However, the power constraint is a minor issue for UEs in SCNs
since they are generally not power-limited due to the close proximity
of a UE and its associated SCN BS. For example, it is recommended
in~\cite{TR36.828} that $P_{\textrm{max}}$ is smaller than the
SCN BS downlink (DL) power by only 1dB, which grants a similar outreach
range of signal transmission for the BS and the UE. Therefore, the
UL power limitation is a minor issue as long as the UE is able to
connect with the BS in the DL. For the sake of tractability, in this
paper, we model $P_{b}$ as (\ref{eq:UL_P_UE}), which has been widely
adopted in the literature {[}3,6-8,11{]}.%
}, e.g., the fractional pathloss compensation (FPC) scheme~\cite{TR36.828}.
Based on this FPC scheme, $P_{b}$ in dBm %
is modeled as%

\noindent
\begin{equation}
P_{b}={P_{0}}+\eta\left({L_{bb}+S_{bb}}\right),\label{eq:UL_P_UE}
\end{equation}
where $P_{0}$ is the target received power at the BS in dBm on the
considered frequency RB, $\eta\in\left(0,1\right]$ is the FPC factor,
$L_{bb}$ has been defined in (\ref{eq:PL_BS2UE}), and $S_{bb}\sim\mathcal{N}\left(0,\sigma_{\textrm{Shad}}^{2}\right)$
has been discussed above.

\textit{The multi-path fading} channel %
from UE $K_{b}$ to the BS of $C_{m}$ is denoted by ${\bf {h}}_{bm}\in\mathbb{C}$%
, where we assume that each UE and each BS are equipped with one omni-directional
antenna. In this paper, we consider a general type of multi-path fading
by assuming that the effective channel gain (in dB) associated with
${\bf {h}}_{bm}$ is defined as $10\log_{10}\left|{\bf {h}}_{bm}\right|^{2}$
and denoted by $H_{bm}$, which follows an i.i.d. distribution with
the PDF of $f_{H}\left(h\right)$. For example, $\left|{\bf {h}}_{bm}\right|^{2}$
can be respectively characterized by an exponential distribution or
a non-central chi-squared distribution in case of Rayleigh fading
or Rician fading~\cite{Book_Proakis}.

\noindent %
\emph{}%

\section{Analysis of the UL Interference Distribution\label{sec:Analysis-of-UL-Interf}}

Based on the definition of RVs discussed in Section~\ref{sec:Network-Model}%
, \emph{the UL received interference power} (in dBm) from UE $K_{b}$
to the BS of $C_{1}$ can be written as%

{\small{}\vspace{-0.4cm}
}{\small \par}

\noindent
\begin{eqnarray}
I_{b} & \hspace{-0.3cm}\overset{(a)}{=}\hspace{-0.3cm} & P_{b}-L_{b1}-S_{b1}+H_{b1}\nonumber \\
 & \hspace{-0.3cm}=\hspace{-0.3cm} & P_{0}+\left(\eta L_{bb}-L_{b1}\right)+\left(\eta S_{bb}-S_{b1}\right)+H_{b1}\nonumber \\
 & \hspace{-0.3cm}\overset{\triangle}{=}\hspace{-0.3cm} & \left(P_{0}+L+S\right)+H_{b1},\nonumber \\
 & \hspace{-0.3cm}\overset{\triangle}{=}\hspace{-0.3cm} & I_{b}^{\left(1\right)}+H_{b1},\label{eq:rx_interf_I1b_UL}
\end{eqnarray}

\noindent where (\ref{eq:UL_P_UE}) is plugged into the step (a) of
(\ref{eq:rx_interf_I1b_UL}), and $L$ and $S$ are defined as $L\overset{\triangle}{=}\left(\eta L_{bb}-L_{b1}\right)$
and $S\overset{\triangle}{=}\left(\eta S_{bb}-S_{b1}\right)$, respectively.
Apparently, $L$ and $S$ are independent RVs. Besides, the first
part of $I_{b}$ is further defined as $I_{b}^{\left(1\right)}\overset{\triangle}{=}\left(P_{0}+L+S\right)$.
Since $S_{bb}$ and $S_{b1}$ $\left(b\in\left\{ 2,\ldots,B\right\} \right)$
are i.i.d. zero-mean Gaussian RVs, it is easy to show that $S$ is
also a Gaussian RV, whose mean and variance are

\noindent
\begin{equation}
\begin{cases}
\mu_{S}=0 & \hspace{-0.3cm}\\
\sigma_{S}^{2}=\left(1+\eta^{2}\right)\sigma_{\textrm{Shad}}^{2} & \hspace{-0.3cm}
\end{cases}.\label{eq:comb_shadowing_mean_and_var}
\end{equation}

From the definition of \emph{$I_{b}$} in~(\ref{eq:rx_interf_I1b_UL}),\emph{
the aggregate interference power} (in mW) from all interfering UEs
to the BS of $C_{1}$ can be written as

\noindent
\begin{equation}
I^{\textrm{mW}}=\sum\limits _{b=2}^{B}{10^{\frac{1}{10}I_{b}}}.\label{eq:rx_interf_UL}
\end{equation}

In the following subsections, we will analyze the distribution of
$I^{\textrm{mW}}$ in three steps:
\begin{itemize}
\item First, we investigate the distribution of $I_{b}^{\left(1\right)}$
shown in (\ref{eq:rx_interf_I1b_UL}) and approximate such distribution
by a Gaussian distribution. The upper-bound of the approximation error
measured by the Kolmogorov\textendash Smirnov (KS)~\cite{KS-distance}
distance is derived in closed-form expressions.
\item Second, we analyze the sum of RVs $\left(I_{b}^{\left(1\right)}+H_{b1}\right)$
shown in (\ref{eq:rx_interf_I1b_UL}) and further approximate the
distribution of $I_{b}$ by another Gaussian distribution. The upper-bound
of the approximation error measured by the KS distance~\cite{KS-distance}
is derived in closed-form expressions.
\item Third, we show that the distribution of $I^{\textrm{mW}}$ can be
well approximated by a power lognormal distribution, i.e., with the
CDF being a power function of the CDF of a lognormal distribution%
.
\end{itemize}

\subsection{The Distribution of $I_{b}^{\left(1\right)}$ in~(\ref{eq:rx_interf_I1b_UL})\label{sub:approx_L+S}}

Considering the complicated mathematical form of $L$ in~(\ref{eq:rx_interf_I1b_UL}),
we can find that the PDF of $L$ is generally not tractable because
$L$ is a RV with respect to $d_{bb}$ and $d_{b1}$, which jointly
depend on the arbitrary shape/size of $R_{b}$ and the arbitrary form
of $f_{Z_{b}}\left(z\right)$. Specifically, for any point $z\in R_{b}$,
$d_{bb}$ and $d_{b1}$ are geometric functions of $z$, and the probability
density of $z$ is $f_{Z_{b}}\left(z\right)$. %
Despite the intractable nature of $L$, we will show that $I_{b}^{\left(1\right)}$
can still be approximated by a Gaussian RV with bounded approximation
errors. In more detail, we investigate an upper bound on the error
in approximating the sum of an arbitrary RV and a Gaussian RV, i.e.,
$\left(L+S\right)$, by another Gaussian RV. To that end, we denote
by $\mu_{L}$ and $\sigma_{L}^{2}$ the mean and variance of $L$
respectively. Moreover, we define two zero-mean RVs as $\tilde{L}=L-\mu_{L}$
and $\tilde{S}=S-\mu_{S}$, respectively. As a result, $I_{b}^{\left(1\right)}$
in~(\ref{eq:rx_interf_I1b_UL}) can be re-written as

\noindent
\begin{equation}
I_{b}^{\left(1\right)}=\ensuremath{\left(\tilde{L}+\tilde{S}\right)+\left(P_{0}+\mu_{L}+\mu_{S}\right)}.\label{eq:Ib_reform_rewritten}
\end{equation}

Next, we approximate $\left(\tilde{L}+\tilde{S}\right)$ by a Gaussian
RV. And it follows that $I_{b}^{\left(1\right)}$ can also be approximated
by the same Gaussian RV with an offset $\left(P_{0}+\mu_{L}+\mu_{S}\right)$.

\subsubsection{The Distribution of $\left(\tilde{L}+\tilde{S}\right)$ in~(\ref{eq:Ib_reform_rewritten})\label{sub:Gauss+L}}

$\,$

For the convenience of mathematical expression, the mean, the variance,
the 3rd moment and the 4th moment of $\tilde{L}$ are denoted by $\mu_{\tilde{L}}=0$,
$\sigma_{\tilde{L}}^{2}=\sigma_{L}^{2}$, $\rho_{\tilde{L}}^{\left(3\right)}$
and $\rho_{\tilde{L}}^{\left(4\right)}$, respectively. Besides, considering
(\ref{eq:comb_shadowing_mean_and_var}), the mean and the variance
of $\tilde{S}$ are denoted by $\mu_{\tilde{S}}=0$, $\sigma_{\tilde{S}}^{2}=\sigma_{S}^{2}$,
respectively. Moreover, we define $\tilde{G}\overset{\triangle}{=}\tilde{L}+\tilde{S}$
and denote the PDF of $\tilde{L}$ by $f_{\tilde{L}}\left(l\right)$.

In order to quantify the approximation error between the distribution
$\tilde{G}$ and its approximate Gaussian distribution, we invoke
the following definition on the Kolmogorov\textendash Smirnov (KS)
distance between two CDFs~\cite{KS-distance}.
\begin{defn}
\label{def:KS_dis}Suppose that the CDFs of RVs $X$ and $\hat{X}$
are $F_{X}\left(x\right)$ and $F_{\hat{X}}\left(x\right)$, respectively.
Then the Kolmogorov\textendash Smirnov (KS) distance between $F_{X}\left(x\right)$
and $F_{\hat{X}}\left(x\right)$ is defined as
\end{defn}
\noindent
\begin{equation}
KS\left(X,Y\right)=\underset{x\in\mathbb{R}}{\sup}\left|F_{X}\left(x\right)-F_{\hat{X}}\left(x\right)\right|.\label{eq:KS_dis}
\end{equation}

The KS distance is a widely used metric to measure the difference
between two CDFs%
. Based on Definition~\ref{def:KS_dis}, we present Theorem~\ref{thm:Gauss+sth=00003DGauss}
in the following to bound the KS distance between the CDF of $\tilde{G}$
and that of the corresponding approximate zero-mean Gaussian RV with
a variance of $\left(\sigma_{L}^{2}+\sigma_{S}^{2}\right)$.
\begin{thm}
\label{thm:Gauss+sth=00003DGauss}Considering the zero-mean RV $\tilde{G}=\tilde{L}+\tilde{S}$
given by~(\ref{eq:Ib_reform_rewritten}), the KS distance between
the CDF of $\tilde{G}$ and that of the corresponding approximate
zero-mean Gaussian RV with a variance of $\left(\sigma_{L}^{2}+\sigma_{S}^{2}\right)$
is bounded by

\noindent
\begin{equation}
\underset{g\in\mathbb{R}}{\sup}\left|F_{\tilde{G}}\left(g\right)-\Phi\left(\frac{g}{\sqrt{\sigma_{L}^{2}+\sigma_{S}^{2}}}\right)\right|\leq\max\left\{ \varepsilon_{1}+\varepsilon_{2},\:\varepsilon_{3}\right\} ,\label{eq:ineq_thm_Gauss}
\end{equation}

\noindent where $F_{\tilde{G}}\left(g\right)$ and $\Phi\left(\cdot\right)$
are respectively the CDF of $\tilde{G}$ and the CDF of the standard
normal distribution, and $\max\left\{ x,y\right\} $ extracts the
larger value between $x$ and $y$.

Furthermore, $\varepsilon_{1}$ in~(\ref{eq:ineq_thm_Gauss}) is
expressed as

\noindent
\begin{equation}
\varepsilon_{1}=\frac{1}{2}\delta_{1}\left(\omega,p\right)\frac{1}{k_{2}^{2}}+\frac{1}{2}\delta_{0}\left(\omega,p,\frac{\left(k_{1}+k_{2}\right)\sqrt{\sigma_{L}^{2}+\sigma_{S}^{2}}}{\sqrt{2}\sigma_{S}}\right)+\frac{1}{2}\delta_{0}\left(\frac{\omega\sqrt{\sigma_{L}^{2}+\sigma_{S}^{2}}}{\sigma_{S}},p,\frac{k_{1}}{\sqrt{2}}\right),\label{eq:epslong_1}
\end{equation}

\noindent where $\omega$, $p$, $k_{1}$ and $k_{2}$ are positive
scalars. Besides, $f_{\tilde{L}}\left(l\right)$ is the PDF of $\tilde{L}$,
and $\delta_{0}\left(\omega,p,k\right)$ and $\delta_{1}\left(\omega,p,k\right)$
are given by

\noindent
\begin{equation}
\begin{cases}
\delta_{0}\left(\omega,p,k\right)=\frac{2}{\sqrt{\pi}\omega}\textrm{erfc}\left(\left(2p+1\right)\omega\right)+\textrm{erfc}\left(\frac{\pi}{2\omega}-k\right) & \hspace{-0.3cm}\\
\delta_{1}\left(\omega,p\right)=\frac{2}{\sqrt{\pi}\omega}\textrm{erfc}\left(\left(2p+1\right)\omega\right)+2 & \hspace{-0.3cm}
\end{cases},\label{eq:def_delta0=0000261}
\end{equation}

\noindent where $\textrm{erfc}\left(\cdot\right)$ is the complementary
error function~\cite{Book_Proakis}.

Moreover, $\varepsilon_{2}$ in~(\ref{eq:ineq_thm_Gauss}) is given
by

\noindent
\begin{equation}
\varepsilon_{2}=\frac{2}{\pi}\sum_{n=1,n\textrm{ odd}}^{2p-1}\left|\upsilon_{n}-\hat{\upsilon}_{n}\right|,\label{eq:epslong_2}
\end{equation}
where $\upsilon_{n}=\frac{1}{n}\exp\left(-\frac{n^{2}\omega^{2}}{2}\right)\varphi_{\tilde{L}}\left(-\frac{n\omega}{\sigma_{S}}\right)$,
$\hat{\upsilon}_{n}=\frac{1}{n}\exp\left(-\frac{n^{2}\omega^{2}}{2}\left(\frac{\sigma_{L}^{2}+\sigma_{S}^{2}}{\sigma_{S}^{2}}\right)\right)$,
and $\varphi_{\tilde{L}}\left(t\right)$ is the characteristic function~\cite{Book_math_tables}
of $f_{\tilde{L}}\left(l\right)$.

Finall, $\varepsilon_{3}$ in~(\ref{eq:ineq_thm_Gauss}) is expressed
as

\noindent
\begin{equation}
\varepsilon_{3}=\frac{1}{k_{1}^{2}}+\frac{1}{2}\textrm{erfc}\left(k_{1}\right).\label{eq:epslong_3}
\end{equation}
\end{thm}
\begin{IEEEproof}
See Appendix~A.
\end{IEEEproof}

Theorem~\ref{thm:Gauss+sth=00003DGauss} is useful to quantify the
maximum error of approximating $\tilde{G}$ by a Gaussian RV%
. From the proof in Appendix~A, it can be seen that
\begin{itemize}
\item $\varepsilon_{1}$ in (\ref{eq:ineq_thm_Gauss}) is caused by the
residual errors from the $p$-truncated Fourier series expansion of
$\textrm{erfc}\left(\cdot\right)$%
, where $p$ is the number of the truncated terms.
\item $\varepsilon_{2}$ in (\ref{eq:ineq_thm_Gauss}) is the major contributor
to the derived upper-bound of the KS distance in Theorem~\ref{thm:Gauss+sth=00003DGauss}.
\item $\varepsilon_{3}$ in (\ref{eq:ineq_thm_Gauss}) measures the asymptotic
difference between $F_{\tilde{G}}\left(g\right)$ and $\Phi\left(\frac{g}{\sqrt{\sigma_{L}^{2}+\sigma_{S}^{2}}}\right)$
for $\left|g\right|\geq k_{1}\sqrt{\sigma_{L}^{2}+\sigma_{S}^{2}}$,
where $k_{1}$ takes a large value to show the asymptotic behavior
of the interested CDFs.
\end{itemize}

To obtain some insights on the typical values of $\varepsilon_{1}$,
$\varepsilon_{2}$ and $\varepsilon_{3}$, we should first discuss
the appropriate choices of the values of $\omega$, $p$, $k_{1}$
and $k_{2}$. To that end, we have the following remarks.

\textbf{Remark 1:} $\omega$ is the fundamental frequency of the Fourier
series expansion of $\textrm{erfc}\left(\cdot\right)$, and thus it
should satisfy the condition that $\frac{1}{\omega}$ is much larger
than the spread of $x$ in $\textrm{erfc}\left(x\right)$~\cite{Book_Proakis},
which can be estimated to be in the range of $\left[-5,5\right]$
because $\left|\textrm{erfc}\left(-\infty\right)-\textrm{erfc}\left(-5\right)\right|=1.54\times10^{-12}\approx0$
and $\left|\textrm{erfc}\left(+\infty\right)-\textrm{erfc}\left(5\right)\right|=1.54\times10^{-12}\approx0$.
Therefore, $\frac{1}{\omega}$ should be much larger than 10 due to
the range of $\left[-5,5\right]$. We propose that a typical value
of $\frac{1}{\omega}$ could be 1000, i.e., $\omega=0.001$.

\textbf{Remark 2:} $p$ is the number of the truncated terms in the
$p$-truncated Fourier series expansion of $\textrm{erfc}\left(\cdot\right)$,
and thus it should be sufficiently large to make the residual error
caused by the truncation of the Fourier series expansion very small.
The part of the residual error that is related to $p$, is expressed
as $\tilde{\delta}\left(\omega,p\right)=\frac{2}{\sqrt{\pi}\omega}\textrm{erfc}\left(\left(2p+1\right)\omega\right)$
in (\ref{eq:def_delta0=0000261}). From this expression, we can see
that $p$ should be larger than $\frac{2}{\omega}$ to make $\tilde{\delta}\left(\omega,p\right)$
sufficiently small, because $\textrm{erfc}\left(\left(2\times\frac{2}{\omega}+1\right)\omega\right)=1.4\times10^{-8}$
when $\omega=0.001$. We propose that a typical value of $p$ could
be $\frac{4}{\omega}$, which equals to 4000 when $\omega=0.001$.
The corresponding $\tilde{\delta}\left(\omega,p\right)$ will then
drop to a very small value of $1.1\times10^{-27}$.

\textbf{Remark 3:} Since $k_{1}$ is related to the asymptotic behavior
of $F_{\tilde{G}}\left(g\right)$ and $\Phi\left(\frac{g}{\sqrt{\sigma_{L}^{2}+\sigma_{S}^{2}}}\right)$
when $\left|g\right|\geq k_{1}\sqrt{\sigma_{L}^{2}+\sigma_{S}^{2}}$,
and thus $k_{1}$ should be sufficiently large to make $\varepsilon_{3}$
a small value. We propose that the typical value of $k_{1}$ should
be at least 100 to achieve an $\varepsilon_{3}$ as small as $1.0\times10^{-4}$,
which corresponds to a small error of $\pm0.01$ percentile. As will
be shown in Section~\ref{sec:Simulaiton-and-Discussion}, $k_{1}$
is set to 500, and we have $\varepsilon_{3}=4\times10^{-6}$, which
is confirmed to be always smaller than $\varepsilon_{2}$ in our numerical
results.

\textbf{Remark 4:} As discussed in Appendix~A, the introduction of
$k_{2}$ is to facilitate the bounding of the integral over $l$ from
$-\infty$ to $+\infty$. From (\ref{eq:epslong_1}), the first term
of (\ref{eq:epslong_1}), i.e., $\frac{1}{2}\delta_{1}\left(\omega,p\right)\frac{1}{k_{2}^{2}}$,
dominates $\varepsilon_{1}$ because $\delta_{0}\left(\omega,p,k\right)$
is a very small value with appropriate choices of $\omega$, $p$,
$k_{1}$ and $k_{2}$. For example, as discussed above, we choose
$\omega=0.001$, $p=4000$, $k_{1}=k_{2}=500$, and $\sigma_{L}=\sigma_{S}$,
then the first term of (\ref{eq:epslong_1}) becomes $4\times10^{-6}$,
while the sum of the rest terms is merely $6.3\times10^{-27}$. Since
a larger $k_{2}$ directly leads to a smaller $\varepsilon_{1}$,
we propose that the typical value of $k_{2}$ should be at least 100
to achieve an $\varepsilon_{1}$ as small as $1.0\times10^{-4}$,
which corresponds to a small error of $\pm0.01$ percentile. As will
be shown in Section~\ref{sec:Simulaiton-and-Discussion}, $k_{2}$
is set to 500, and we have $\varepsilon_{1}=4\times10^{-6}$, which
is confirmed to be always smaller than $\varepsilon_{2}$ in our numerical
results.

Considering the expressions of $\varepsilon_{1}$ and $\varepsilon_{3}$
shown in (\ref{eq:epslong_1}) and (\ref{eq:epslong_3}), respectively,
we investigate a special case of $k_{1}=k_{2}$ and it is straightforward
to get $\varepsilon_{1}>\varepsilon_{3}$. Hence, we propose the following
corollary to simplify Theorem~\ref{thm:Gauss+sth=00003DGauss}.
\begin{cor}
\label{cor:from_main_thm}When $k_{1}=k_{2}$, the KS distance between
the CDF of $\tilde{G}$ and that of the corresponding approximate
zero-mean Gaussian RV with a variance of $\left(\sigma_{L}^{2}+\sigma_{S}^{2}\right)$
is bounded by

\noindent
\begin{equation}
\underset{g\in\mathbb{R}}{\sup}\left|F_{\tilde{G}}\left(g\right)-\Phi\left(\frac{g}{\sqrt{\sigma_{L}^{2}+\sigma_{S}^{2}}}\right)\right|\leq\varepsilon_{1}+\varepsilon_{2}.\label{eq:ineq_thm_Gauss_corollary}
\end{equation}

\end{cor}

Note that compared with (\ref{eq:ineq_thm_Gauss}), $\varepsilon_{3}$
does not exist in the right-hand side of (\ref{eq:ineq_thm_Gauss_corollary}),
which largely simplifies our analysis and discussion. Therefore, in
the sequel we will only consider (\ref{eq:ineq_thm_Gauss_corollary})
to quantify the maximum error of approximating $\tilde{G}$ by a Gaussian
RV.

With the discussed typical values of $\omega$, $p$, $k_{1}$ and
$k_{2}$, we can see that $\varepsilon_{1}$ in (\ref{eq:ineq_thm_Gauss_corollary})
can be controlled to be as small as $4\times10^{-6}$, which leaves
$\varepsilon_{2}$ as the major contributor to the derived upper-bound
of the KS distance. We will briefly discuss the calculation of $\varepsilon_{2}$
in the next subsection. Note that further optimization of $k_{1}$
and $k_{2}$ to reduce $\varepsilon_{3}$ and $\varepsilon_{1}$ even
below $4\times10^{-6}$ is possible. However, such optimization has
a marginal impact on the derived upper bound of the approximation
error because $\varepsilon_{2}$ is independent of $k_{1}$ and $k_{2}$.

With Theorem~\ref{thm:Gauss+sth=00003DGauss} and Corollary~\ref{cor:from_main_thm}
characterizing the upper bound of the approximation error, we propose
to approximate $I_{b}^{\left(1\right)}$ in~(\ref{eq:Ib_reform_rewritten})
by a Gaussian RV $G_{b}$, whose mean and variance can be computed
by

\noindent
\begin{equation}
\begin{cases}
\mu_{G_{b}}=P_{0}+\mu_{L}+\mu_{S} & \hspace{-0.3cm}\\
\ensuremath{\sigma_{G_{b}}^{2}}=\sigma_{L}^{2}+\sigma_{S}^{2} & \hspace{-0.3cm}
\end{cases}.\label{eq:approx_LxLN_mean_and_var}
\end{equation}

\subsubsection{The Calculation of $\varepsilon_{2}$ in (\ref{eq:epslong_2})\label{sub:calc_epslong2}}

$\,$

For each small cell $C_{b}$, considering the definition of $L$ and
$\tilde{L}$ respectively presented in~(\ref{eq:rx_interf_I1b_UL})
and~(\ref{eq:Ib_reform_rewritten}), we can evaluate $\varphi_{\tilde{L}}\left(-\frac{n\omega}{\sigma_{S}}\right)$
for each small cell $C_{b}$ as

\noindent
\begin{eqnarray}
\varphi_{\tilde{L}}\left(-\frac{n\omega}{\sigma_{S}}\right)\hspace{-0.3cm} & = & \hspace{-0.3cm}\int_{-\infty}^{+\infty}\exp\left(-\textrm{j}\frac{n\omega l}{\sigma_{S}}\right)f_{\tilde{L}}\left(l\right)dl\nonumber \\
\hspace{-0.3cm} & = & \hspace{-0.3cm}\int_{R_{b}}\exp\left(-\textrm{j}n\omega\frac{L\left(z\right)-\mu_{L}}{\sigma_{S}}\right)f_{Z_{b}}\left(z\right)dz\nonumber \\
\hspace{-0.3cm} & \overset{\left(a\right)}{=} & \hspace{-0.3cm}\int_{R_{b}}\exp\left(-\textrm{j}n\omega\frac{\eta L_{bb}\left(z\right)-L_{b1}\left(z\right)-\mu_{L}}{\sigma_{S}}\right)f_{Z_{b}}\left(z\right)dz,\label{eq:CF_numerical_integral}
\end{eqnarray}

\noindent where (\ref{eq:PL_BS2UE}) should be inserted into the step
(a) of (\ref{eq:CF_numerical_integral}) and $\mu_{L}$ is the mean
of $L$.

With the result of $\varphi_{\tilde{L}}\left(-\frac{n\omega}{\sigma_{S}}\right)$,
we can then compute $\varepsilon_{2}$ according to its definition
in (\ref{eq:epslong_2}). %

\subsection{The Distribution of $I_{b}$ in~(\ref{eq:rx_interf_I1b_UL})\label{sub:approx_LNxMPFad}}

Having approximated $I_{b}^{\left(1\right)}$ by a Gaussian RV $G_{b}$,
we can approximate~(\ref{eq:rx_interf_I1b_UL}) as

\noindent
\begin{equation}
I_{b}\approx G_{b}+H_{b1}.\label{eq:Ib_reform}
\end{equation}

\noindent It is interesting to note that, similar to $I_{b}^{\left(1\right)}$
in (\ref{eq:rx_interf_I1b_UL}), the approximate expression of $I_{b}$
in (\ref{eq:Ib_reform}) also contains a Gaussian RV $G_{b}$ and
an arbitrary RV $H_{b1}$ with the PDF of $f_{H}\left(h\right)$%
.

Based on the above observation, we propose to reuse Theorem~\ref{thm:Gauss+sth=00003DGauss}
and Corollary~\ref{cor:from_main_thm} to quantify the error in approximating
$I_{b}$ in (\ref{eq:Ib_reform}) by an another Gaussian RV. To that
end, similar to (\ref{eq:Ib_reform_rewritten}), we define two zero-mean
RVs as $\tilde{G}_{b}=G_{b}-\mu_{G_{b}}$ and $\tilde{H}_{b1}=H_{b1}-\mu_{H_{b1}}$,
where $\mu_{G_{b}}$ and $\mu_{H_{b1}}$ are the means of $G_{b}$
and $H_{b1}$, respectively. Besides, the variance of $G_{b}$ and
$H_{b1}$ are denoted by $\sigma_{G_{b}}^{2}$ and $\sigma_{H_{b1}}^{2}$,
respectively. As a result, we can re-formulate (\ref{eq:Ib_reform})
as

\noindent
\begin{equation}
I_{b}\approx\left(\tilde{G}_{b}+\tilde{H}_{b1}\right)+\left(\mu_{G_{b}}+\mu_{H_{b1}}\right).\label{eq:Ib_further_reform}
\end{equation}

Next, we approximate $\left(\tilde{G}_{b}+\tilde{H}_{b1}\right)$
by a Gaussian RV. And it follows that $I_{b}$ can also be approximated
by the same Gaussian RV with an offset $\left(\mu_{G_{b}}+\mu_{H_{b1}}\right)$.

\subsubsection{The Distribution of $\left(\tilde{G}_{b}+\tilde{H}_{b1}\right)$
in (\ref{eq:Ib_further_reform})\label{sub:Gauss+MPFad}}

$\,$

We propose to approximate $\left(\tilde{G}_{b}+\tilde{H}_{b1}\right)$
as a zero-mean Gaussian RV with a variance of $\sigma_{G_{b}}^{2}+\sigma_{H_{b1}}^{2}$,
then from%
{} Corollary~\ref{cor:from_main_thm}, the error measured by the KS
distance between the actual CDF and the approximate CDF can be upper-bounded
by $\varepsilon'_{1}+\varepsilon'_{2}$, where $\varepsilon'_{1}$
and $\varepsilon'_{2}$ are respectively computed using (\ref{eq:epslong_1})
and (\ref{eq:epslong_2}) with the following RV changes,%

\noindent
\begin{equation}
\begin{cases}
\varepsilon'_{1}=\left[\varepsilon_{1}\left|S\rightarrow G_{b},\tilde{S}\rightarrow\tilde{G}_{b},L\rightarrow H_{b1},\tilde{L}\rightarrow\tilde{H}_{b1}\right.\right] & \hspace{-0.3cm}\\
\varepsilon'_{2}=\left[\varepsilon_{2}\left|S\rightarrow G_{b},\tilde{S}\rightarrow\tilde{G}_{b},L\rightarrow H_{b1},\tilde{L}\rightarrow\tilde{H}_{b1}\right.\right] & \hspace{-0.3cm}
\end{cases},\label{eq:epslong_prime}
\end{equation}

\noindent where $X\rightarrow Y$ denotes the RV change of replacing
$X$ with $Y$.

\subsubsection{The approximate PDF and CDF of $I_{b}$\label{sub:The-PDF-CDF-of-Ib}}

$\,$

With the approximation of $\left(\tilde{G}_{b}+\tilde{H}_{b1}\right)$
as a zero-mean Gaussian RV, we propose to approximate $I_{b}$ in
(\ref{eq:Ib_further_reform}) as another Gaussian RV $Q_{b}$, whose
mean and variance are

\noindent
\begin{equation}
\begin{cases}
\mu_{Q_{b}}=\mu_{G_{b}}+\mu_{H_{b1}} & \hspace{-0.3cm}\\
\ensuremath{\sigma_{Q_{b}}^{2}}=\sigma_{G_{b}}^{2}+\sigma_{H_{b1}}^{2} & \hspace{-0.3cm}
\end{cases}.\label{eq:approx_LxLNxEXP_mean_and_var}
\end{equation}

According to Theorem~\ref{thm:Gauss+sth=00003DGauss} and Corollary~\ref{cor:from_main_thm},
and from (\ref{eq:epslong_1}), (\ref{eq:epslong_2}) and (\ref{eq:epslong_prime}),
the total error of approximating $I_{b}$ as $Q_{b}$, measured by
the KS distance between the CDF of $I_{b}$ and that of $Q_{b}$,
can be upper-bounded by $\varepsilon$ as

\noindent
\begin{equation}
\varepsilon=\left(\varepsilon_{1}+\varepsilon_{2}\right)+\left(\varepsilon'_{1}+\varepsilon'_{2}\right),\label{eq:sum_epslong}
\end{equation}

\noindent where $\left(\varepsilon_{1}+\varepsilon_{2}\right)$ is
the upper bound on the error associated with the approximation of
$I_{b}^{\left(1\right)}$ in~(\ref{eq:Ib_reform_rewritten}) as a
Gaussian RV $G_{b}$, and $\left(\varepsilon'_{1}+\varepsilon'_{2}\right)$
is the upper bound on the error associated with the approximation
of $G_{b}+H_{b1}$ in~(\ref{eq:Ib_reform}) as a Gaussian RV $Q_{b}$.

\subsection{The Distribution of $I^{\textrm{mW}}$ in~(\ref{eq:rx_interf_UL})\label{sub:approx_I_mw}}

The study on the approximate distribution of the sum of multiple independent
lognormal RVs has been going on for more than five decades~\cite{Approx_sumLN}-\hspace{-0.03cm}\cite{power_LN_approx_GC}%
. According to~\cite{Approx_sumLN}-\hspace{-0.03cm}\cite{Opt_approx_sumLN}%
, the sum of multiple independent lognormal RVs can be well approximated
by another lognormal RV. However, some recent studies~\cite{power_LN_approx_TVT}-\hspace{-0.03cm}\cite{power_LN_approx_GC}%
{} concluded that the sum of multiple independent lognormal RVs is better
approximated by a power lognormal RV, i.e., with the CDF being a power
function of $\Phi\left(\cdot\right)$. In this paper, we adopt the
power lognormal approximation for $I^{\textrm{mW}}$%
, which will be explained in the following.

In our case, %
since each $I_{b},b\in\left\{ 2,\dots,B\right\} $ is approximated
by the Gaussian RV $Q_{b}$, the sum of $10^{\frac{1}{{10}}Q_{b}}$
can be well approximated by a power lognormal RV~\cite{power_LN_approx_TVT}-\hspace{-0.03cm}\cite{power_LN_approx_GC}%
{} expressed as $\hat{I}^{\textrm{mW}}=10^{\frac{1}{{10}}Q}$, where
the PDF and CDF of $Q$ can be respectively written as~\cite{power_LN_approx_TVT}

\noindent
\begin{equation}
\begin{cases}
\textrm{PDF of }Q:\:{f_{Q}}\left(q\right)=\lambda\Phi^{\lambda-1}\left(\frac{q-{\mu_{Q}}}{\sigma_{Q}}\right)\frac{1}{\sqrt{2\pi\ensuremath{\sigma_{Q}^{2}}}}\exp\left\{ {-\frac{{\left({q-{\mu_{Q}}}\right)^{2}}}{{2\ensuremath{\sigma_{Q}^{2}}}}}\right\}  & \hspace{-0.3cm}\\
\textrm{CDF of }Q:\:{F_{Q}}\left(q\right)=\Phi^{\lambda}\left(\frac{q-{\mu_{Q}}}{\sigma_{Q}}\right) & \hspace{-0.3cm}
\end{cases},\label{eq:PDF_CDF_powerLN_Q}
\end{equation}

\noindent %

\noindent where the parameters $\lambda$, $\mu_{Q}$ and $\sigma_{Q}$
are obtained from $\left\{ \mu_{Q_{b}}\right\} $ and $\left\{ \ensuremath{\sigma_{Q_{b}}^{2}}\right\} $.
The method to accomplish such task has been well addressed in~\cite{power_LN_approx_TVT}-\hspace{-0.03cm}\cite{power_LN_approx_GC}%
.%
{} In Appendix~B, we provide an example to obtain $\lambda$, $\mu_{Q}$
and $\sigma_{Q}$ based on~\cite{Approx_sumLN},~\cite{para_deci_power_LN_approx}
and~\cite{power_LN_approx_GC}.

As a result of (\ref{eq:PDF_CDF_powerLN_Q}), the PDF and CDF of $\hat{I}^{\textrm{mW}}$
can be respectively written as

\noindent
\begin{equation}
\begin{cases}
\textrm{PDF of }\hat{I}^{\textrm{mW}}:\:{f_{\hat{I}^{\textrm{mW}}}}\left(v\right)=\lambda\Phi^{\lambda-1}\left(\frac{\zeta\ln v-{\mu_{Q}}}{\sigma_{Q}}\right)\frac{\zeta}{v\sqrt{2\pi\ensuremath{\sigma_{Q}^{2}}}}\exp\left\{ {-\frac{{\left({\zeta\ln v-{\mu_{Q}}}\right)^{2}}}{{2\ensuremath{\sigma_{Q}^{2}}}}}\right\}  & \hspace{-0.3cm}\\
\textrm{CDF of }\hat{I}^{\textrm{mW}}:\:{F_{\hat{I}^{\textrm{mW}}}}\left(v\right)=\Phi^{\lambda}\left(\frac{\zeta\ln v-{\mu_{Q}}}{\sigma_{Q}}\right) & \hspace{-0.3cm}
\end{cases},\label{eq:PDF_CDF_agg_Interf}
\end{equation}

\noindent %

\noindent where $\zeta=\frac{10}{\ln10}$ is a scalar caused by the
variable change from $10\log_{10}v$ to $\ln v$.

Finally, we propose that the distribution of $I^{\textrm{mW}}$ can
be approximated by that of $\hat{I}^{\textrm{mW}}$ shown in (\ref{eq:PDF_CDF_agg_Interf}).
Note that in this step of approximation, the approximation error is
dependent on the adopted approximate distribution of the sum of multiple
independent lognormal RVs. We will study such approximation error
in our future work. Note that some recent studies~\cite{power_LN_approx_TVT}-\hspace{-0.03cm}\cite{power_LN_approx_GC}%
{} have shown that the error associated with the power lognormal approximation
is reasonably small and good enough for practical use.

\subsection{Summary of the Proposed Analysis of the UL Interference Distribution\label{sub:Summary_approx_analysis}}

To sum up, in the following, we highlight the main steps in our proposed
microscopic analysis of the UL interference distribution. First, for
each $b\in\left\{ 2,\ldots,B\right\} $, we use (\ref{eq:epslong_1})
and (\ref{eq:epslong_2}) to check $\left(\varepsilon_{1}+\varepsilon_{2}\right)$
associated with the approximation of $I_{b}^{\left(1\right)}$ in~(\ref{eq:Ib_reform_rewritten})
as a Gaussian RV $G_{b}$. %
Second, for each $b\in\left\{ 2,\ldots,B\right\} $, we use (\ref{eq:epslong_prime})
to check $\left(\varepsilon'_{1}+\varepsilon'_{2}\right)$ associated
with the approximation of $G_{b}+H_{b1}$ in~(\ref{eq:Ib_reform})
as a Gaussian RV $Q_{b}$. The upper bound of the total approximation
error of the above two steps is further obtained from (\ref{eq:sum_epslong}),
without any requirement on (i) the shape and/or size of cell coverage
areas, (ii) the uniformity of UE distribution, and (iii) the type
of multi-path fading. Finally, we approximate $I^{\textrm{mW}}$ in~(\ref{eq:rx_interf_UL})
by a power lognormal RV $\hat{I}^{\textrm{mW}}=10^{\frac{1}{{10}}Q}$,
where the PDF and the CDF of $Q$ is expressed as (\ref{eq:PDF_CDF_powerLN_Q})
with the parameters $\lambda$, $\mu_{Q}$ and $\sigma_{Q}^{2}$ obtained
from, e.g., Appendix~B.

\section{Simulation and Discussion\label{sec:Simulaiton-and-Discussion}}

In order to validate the approximation from the proposed microscopic
analysis of the UL interference, we conduct simulations considering
two types of scenarios, i.e., one with a single interfering cell and
the other with multiple interfering cells. As discussed in Remarks~1\textasciitilde{}4
in Subsection~\ref{sub:Gauss+L}, the parameters for evaluating Theorem~\ref{thm:Gauss+sth=00003DGauss}
and Corollary~\ref{cor:from_main_thm} are set to: $\omega=0.001$,
$p=4000$, $k_{1}=k_{2}=500$. %
According to the 3GPP standards~\cite{TR36.828}, the system parameters
are set to: $A=103.8$, $\alpha=20.9$, $P_{0}=-76$\ dBm, $\eta=0.8$,
and $\sigma_{S}=10$\ dB. Besides, the minimum BS-to-UE distance
is assumed to be 0.005\ km~\cite{TR36.828}.

\subsection{The Scenario with a Single Interfering Cell\label{sub:The-Scenario-single-cell}}

In this scenario, the number of BSs $B$ is set to 2, and UEs in small
cell $C_{2}$ are assumed to be uniformly distributed and non-uniformly
distributed in the coverage area $R_{2}$, respectively.

\subsubsection{Uniformly Distributed UEs\label{sub:B2_UEUnifDis}}

$\,$

\begin{figure}[H]
\noindent \begin{centering}
\vspace{-0.5cm}
\includegraphics[width=8cm]{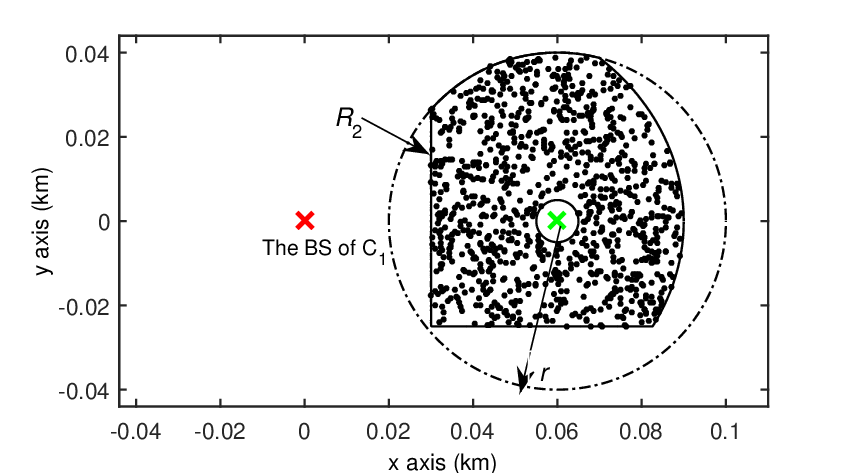}\renewcommand{\figurename}{Fig.}\protect\caption{\label{fig:R2_bread_UEUnifDis}Illustration of the coverage area $R_{2}$
($r=0.04\,\textrm{km}$, uniformly distributed UEs).}

\par\end{centering}

\vspace{-0.5cm}
\end{figure}

In this case, we consider %
uniformly distributed UEs, as shown in Fig.~\ref{fig:R2_bread_UEUnifDis}.
The x-markers indicate BS locations where the BS location of $C_{1}$
has been explicitly pointed out. The dash-dot line indicates a reference
disk to illustrate the reference size of small cell $C_{2}$. The
radius of such a reference circle is denoted by $r$, and the distance
between the BS of $C_{1}$ and the BS of $C_{2}$, i.e., $D_{2}$,
is assumed to be $1.5r$. Note that $1.5r$ is just an example of
$D_{2}$ and the specific value of $D_{2}$ has no impact on the procedure
of our analysis. In our simulations, the values of $r$ (in km) are
set to 0.01, 0.02 and 0.04, respectively~\cite{Tutor_smallcell}.
In this scenario, the interfering UE $K_{2}$ is uniformly distributed
in an irregularly shaped coverage area $R_{2}$, as shown by the area
outlined by the solid line in Fig.~\ref{fig:R2_bread_UEUnifDis}.
The shape of $R_{2}$ is the intersection of a square, a circle and
an ellipse, which has a very complicated generation function. Examples
of the possible positions of $K_{2}$ within $R_{2}$ are shown as
dots in Fig.~\ref{fig:R2_bread_UEUnifDis}.

Despite the complicated shape of $R_{2}$, our proposed microscopic
analysis of the UL interference distribution still can be applied.
Specifically, from Corollary~\ref{cor:from_main_thm}, (\ref{eq:epslong_1}),
(\ref{eq:epslong_2}), and (\ref{eq:epslong_prime}), we can examine
the validity of approximating $I_{2}$ as a Gaussian RV by checking
the corresponding $\varepsilon$ given by (\ref{eq:sum_epslong}).
The results of $\varepsilon$ are tabulated for various values of
$r$ with the assumption of \emph{Rayleigh fading} in Table~\ref{tab:results_bread_UEUnifDis_Rayleigh}.
Moreover, $\varepsilon$ is broken into $\left(\varepsilon_{1}+\varepsilon_{2}\right)$
and $\left(\varepsilon'_{1}+\varepsilon'_{2}\right)$, with the marginal
contribution of $\varepsilon_{1}$ and $\varepsilon'_{1}$ shown in
Table~\ref{tab:results_bread_UEUnifDis_Rayleigh} as well. From this
table, we can observe that $\varepsilon_{2}$ and $\varepsilon'_{2}$
are about $100$ times larger than $\varepsilon_{1}$ and $\varepsilon'_{1}$,
indicating the dominance of $\varepsilon_{2}$ and $\varepsilon'_{2}$
in the result of $\varepsilon$. For all the investigated values of
$r$, the values of $\varepsilon$ are below 0.01%
. %
Consequently, the approximation of $I_{2}$ as a Gaussian RV should
be tight. Note that $r=0.01$ and $r=0.02$ correspond to the typical
network configurations for future dense and ultra-dense SCNs~\cite{Tutor_smallcell},
which shows that our proposed microscopic analysis of the UL interference
distribution can be readily used to study future dense and ultra-dense
SCNs. Also note that the upper bound $\varepsilon$ is much tighter
than that presented in our previous work~\cite{UL_interference_approx_GC}.
Specifically, when $r=0.04$, the upper bound on the approximation
error in~\cite{UL_interference_approx_GC} is around $1.84\times10^{-2}$,
while in Table~\ref{tab:results_bread_UEUnifDis_Rayleigh} $\varepsilon$
is only $4.6\times10^{-3}$, which is a significant improvement compared
with our previous result in~\cite{UL_interference_approx_GC}.

\begin{table}[H]
\begin{centering}
{\small{}\protect\caption{\label{tab:results_bread_UEUnifDis_Rayleigh}Approximation errors
of the proposed analysis \protect \\
($B=2$, uniformly distributed UEs, Rayleigh fading).}
}
\par\end{centering}{\small \par}

{\small{}\vspace{-0.4cm}
}{\small \par}

\begin{centering}
{\small{}}%
\begin{tabular}{|l|l|l|l|l|l|l|l|l|}
\hline
{\small{}$r$} & {\small{}$\varepsilon$} & {\small{}$\varepsilon_{1}+\varepsilon_{2}$} & {\small{}$\varepsilon_{1}$} & {\small{}$\varepsilon'_{1}+\varepsilon'_{2}$} & {\small{}$\varepsilon'_{1}$} & {\small{}$\mu_{Q_{2}}$} & {\small{}$\sigma_{Q_{2}}^{2}$} & {\small{}Actual error}\tabularnewline
\hline
\hline
{\small{}$0.01$} & {\small{}$4.9\times10^{-3}$} & {\small{}$3.5\times10^{-4}$} & {\small{}$4\times10^{-6}$} & {\small{}$4.6\times10^{-3}$} & {\small{}$4\times10^{-6}$} & {\small{}-97.1} & {\small{}205.3} & {\small{}$1.9\times10^{-3}$}\tabularnewline
\hline
{\small{}$0.02$} & {\small{}$5.0\times10^{-3}$} & {\small{}$5.2\times10^{-4}$} & {\small{}$4\times10^{-6}$} & {\small{}$4.5\times10^{-3}$} & {\small{}$4\times10^{-6}$} & {\small{}-99.7} & {\small{}207.7} & {\small{}$1.9\times10^{-3}$}\tabularnewline
\hline
{\small{}$0.04$} & {\small{}$4.6\times10^{-3}$} & {\small{}$1.1\times10^{-4}$} & {\small{}$4\times10^{-6}$} & {\small{}$4.5\times10^{-3}$} & {\small{}$4\times10^{-6}$} & {\small{}-101.5} & {\small{}211.4} & {\small{}$1.8\times10^{-3}$}\tabularnewline
\hline
\end{tabular}
\par\end{centering}{\small \par}

\vspace{-0.5cm}
\end{table}
 %

\begin{figure}[H]
\noindent \begin{centering}
\vspace{-0.5cm}
\includegraphics[width=8cm]{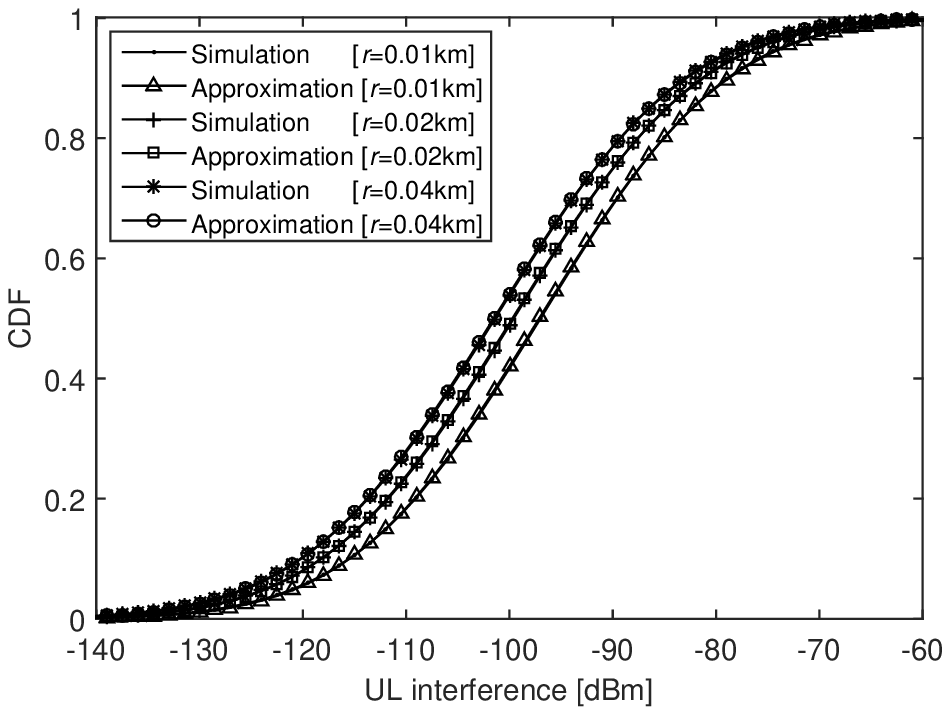}\renewcommand{\figurename}{Fig.}\protect\caption{\label{fig:UL_interf_approx_irregR2_UEUnifDis_Rayleigh}The simulation
and the approximation of the CDF of $I_{2}$ ($B=2$, uniformly distributed
UEs, Rayleigh fading).}

\par\end{centering}

\vspace{-0.5cm}
\end{figure}

To further verify the accuracy of the proposed approximation, we plot
the simulation results of the CDF of $I_{2}$ and the analytical results
of the approximate Gaussian CDF according to (\ref{eq:approx_LxLNxEXP_mean_and_var})
for the considered $R_{2}$ in Fig.~\ref{fig:UL_interf_approx_irregR2_UEUnifDis_Rayleigh}.
The numerical results of $\mu_{Q_{2}}$ and $\sigma_{Q_{2}}^{2}$
are also listed in Table~\ref{tab:results_bread_UEUnifDis_Rayleigh}.
As can be seen from Fig.~\ref{fig:UL_interf_approx_irregR2_UEUnifDis_Rayleigh},
the proposed Gaussian approximation of $I_{2}$ is very tight for
the considered $R_{2}$ with such an irregular shape. Note that the
actual errors between the simulation and the approximation of the
CDF of $I_{2}$ are also shown in Table~\ref{tab:results_bread_UEUnifDis_Rayleigh},
the results of which establish the validity of the derived upper bound
on the approximation error.

As discussed above, Table~\ref{tab:results_bread_UEUnifDis_Rayleigh}
is obtained with the assumption of Rayleigh fading. In the following,
we will check the approximation errors for Rician fading and another
case without multi-path fading. According to~\cite{Book_Proakis},
a ratio between the power in the%
{} line-of-sight (LOS) path and the power in the other scattered paths
should be defined for Rician fading. Such ratio is denoted by $\Gamma$
in our paper. Note that Rician fading will degrade to Rayleigh fading
when $\Gamma=0$. For various values of $r$ under the assumption
of Rician fading with $\Gamma=10$, the results of $\varepsilon$
are tabulated in Table~\ref{tab:results_bread_UEUnifDis_Rician}.
Comparing Table~\ref{tab:results_bread_UEUnifDis_Rician} with Table~\ref{tab:results_bread_UEUnifDis_Rayleigh},
we can see that the difference lies in the values of $\left(\varepsilon'_{1}+\varepsilon'_{2}\right)$
because it solely depends on the assumption of multi-path fading as
discussed in Subsection~\ref{sub:Gauss+MPFad}. Note that the errors
caused by the consideration of Rician fading is actually smaller than
those of Rayleigh fading, because Rician fading incurs less randomness
than Rayleigh fading due to the dominant LOS path component in the
multi-path fading. Such reduction of randomness can be further observed
in an extreme case of $\Gamma=+\infty$, where a deterministic LOS
path completely dominates the multi-path fading. The results of $\varepsilon$
for such extreme case with $\Gamma=+\infty$, i.e., no multi-path
fading, are exhibited in Table~\ref{tab:results_bread_UEUnifDis_noMPFad},
where $\varepsilon'_{1}$ and $\varepsilon'_{2}$ are all zeros because
the approximation step addressed in Subsection~\ref{sub:approx_LNxMPFad}
is skipped, rendering $H_{b1}\equiv0\,\textrm{dB}$.

For all the investigated values of $r$, the values of $\varepsilon$
are smaller than 0.001 in both Table~\ref{tab:results_bread_UEUnifDis_Rician}
and Table~\ref{tab:results_bread_UEUnifDis_noMPFad}, indicating
that the approximation of $I_{2}$ as a Gaussian RV should be even
tighter than that observed in Fig.~\ref{fig:UL_interf_approx_irregR2_UEUnifDis_Rayleigh}.
For brevity, we omit these figures. %
Note that the actual errors between the simulation and the approximation
of the CDF of $I_{2}$ are provided in Table~\ref{tab:results_bread_UEUnifDis_Rician}
and Table~\ref{tab:results_bread_UEUnifDis_noMPFad}, the results
of which also establish the validity of the derived upper bound $\varepsilon$.

\begin{table}[H]
\begin{centering}
{\small{}\protect\caption{\label{tab:results_bread_UEUnifDis_Rician}Approximation errors of
the proposed analysis \protect \\
($B=2$, uniformly distributed UEs, Rician fading with $\Gamma=10$).}
}
\par\end{centering}{\small \par}

{\small{}\vspace{-0.4cm}
}{\small \par}

\begin{centering}
{\small{}}%
\begin{tabular}{|l|l|l|l|l|l|l|l|l|}
\hline
{\small{}$r$} & {\small{}$\varepsilon$} & {\small{}$\varepsilon_{1}+\varepsilon_{2}$} & {\small{}$\varepsilon_{1}$} & {\small{}$\varepsilon'_{1}+\varepsilon'_{2}$} & {\small{}$\varepsilon'_{1}$} & {\small{}$\mu_{Q_{2}}$} & {\small{}$\sigma_{Q_{2}}^{2}$} & {\small{}Actual error}\tabularnewline
\hline
\hline
{\small{}$0.01$} & {\small{}$5.4\times10^{-4}$} & {\small{}$3.5\times10^{-4}$} & {\small{}$4\times10^{-6}$} & {\small{}$1.9\times10^{-4}$} & {\small{}$4\times10^{-6}$} & {\small{}-95.0} & {\small{}178.2} & {\small{}$2.0\times10^{-4}$}\tabularnewline
\hline
{\small{}$0.02$} & {\small{}$7.0\times10^{-4}$} & {\small{}$5.2\times10^{-4}$} & {\small{}$4\times10^{-6}$} & {\small{}$1.9\times10^{-4}$} & {\small{}$4\times10^{-6}$} & {\small{}-97.6} & {\small{}180.8} & {\small{}$4.8\times10^{-4}$}\tabularnewline
\hline
{\small{}$0.04$} & {\small{}$3.0\times10^{-4}$} & {\small{}$1.1\times10^{-4}$} & {\small{}$4\times10^{-6}$} & {\small{}$1.9\times10^{-4}$} & {\small{}$4\times10^{-6}$} & {\small{}-99.4} & {\small{}184.5} & {\small{}$1.9\times10^{-4}$}\tabularnewline
\hline
\end{tabular}
\par\end{centering}{\small \par}

\vspace{-1cm}
\end{table}

\begin{table}[H]
\begin{centering}
{\small{}\protect\caption{\label{tab:results_bread_UEUnifDis_noMPFad}Approximation errors of
the proposed analysis \protect \\
($B=2$, uniformly distributed UEs, no multi-path fading).}
}
\par\end{centering}{\small \par}

{\small{}\vspace{-0.4cm}
}{\small \par}

\begin{centering}
{\small{}}%
\begin{tabular}{|l|l|l|l|l|l|l|l|l|}
\hline
{\small{}$r$} & {\small{}$\varepsilon$} & {\small{}$\varepsilon_{1}+\varepsilon_{2}$} & {\small{}$\varepsilon_{1}$} & {\small{}$\varepsilon'_{1}+\varepsilon'_{2}$} & {\small{}$\varepsilon'_{1}$} & {\small{}$\mu_{Q_{2}}$} & {\small{}$\sigma_{Q_{2}}^{2}$} & {\small{}Actual error}\tabularnewline
\hline
\hline
{\small{}$0.01$} & {\small{}$3.5\times10^{-4}$} & {\small{}$3.5\times10^{-4}$} & {\small{}$4\times10^{-6}$} & {\small{}0} & {\small{}0} & {\small{}-94.6} & {\small{}174.2} & {\small{}$1.3\times10^{-4}$}\tabularnewline
\hline
{\small{}$0.02$} & {\small{}$5.2\times10^{-4}$} & {\small{}$5.2\times10^{-4}$} & {\small{}$4\times10^{-6}$} & {\small{}0} & {\small{}0} & {\small{}-97.2} & {\small{}176.7} & {\small{}$2.4\times10^{-4}$}\tabularnewline
\hline
{\small{}$0.04$} & {\small{}$1.1\times10^{-4}$} & {\small{}$1.1\times10^{-4}$} & {\small{}$4\times10^{-6}$} & {\small{}0} & {\small{}0} & {\small{}-99.0} & {\small{}180.5} & {\small{}$0.6\times10^{-4}$}\tabularnewline
\hline
\end{tabular}
\par\end{centering}{\small \par}

\vspace{-0.5cm}
\end{table}

\subsubsection{Non-Uniformly Distributed UEs\label{sub:B2_UENonUnifDis}}

$\,$

\begin{figure}[H]
\noindent \begin{centering}
\vspace{-0.5cm}
\includegraphics[width=8cm]{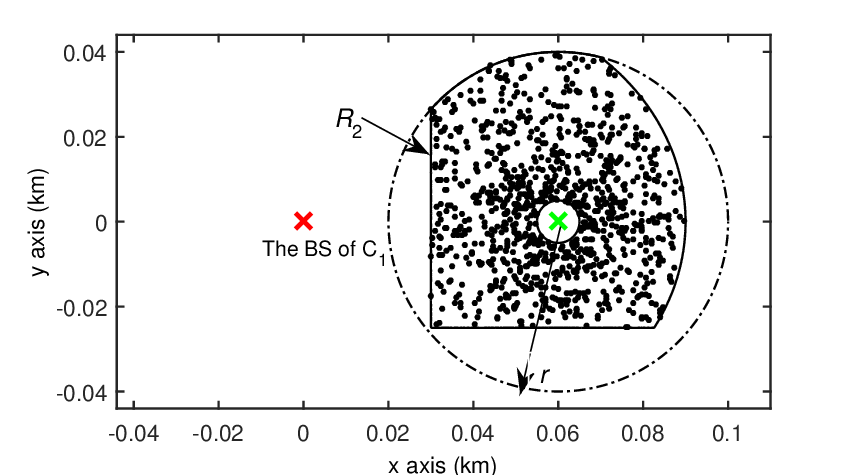}\renewcommand{\figurename}{Fig.}\protect\caption{\label{fig:R2_bread_UENonUnifDis}Illustration of the coverage area
$R_{2}$ ($r=0.04\,\textrm{km}$, non-uniformly distributed UEs).}

\par\end{centering}

\vspace{-0.5cm}
\end{figure}

In this subsection, we further investigate the scenario discussed
in Subsection~\ref{sub:B2_UEUnifDis}. Different from previous assumption,
here we consider that the interfering UE $K_{2}$ is no longer uniformly
distributed in $R_{2}$. In this subsection, we consider a UE distribution
function expressed as $f_{Z_{2}}\left(z\right)=\frac{W}{\rho},z\in R_{2}$,
where $\rho$ is the radial coordinate of $z$ in the polar coordinate
system, the origin of which is placed at the position of the BS of
$C_{2}$. Besides, $W$ is a normalization constant to make $\int_{R_{2}}f_{Z_{2}}\left(z\right)dz=1$.
In the considered non-uniform UE distribution, UEs are more likely
to locate in the close vicinity of the BS of $C_{2}$, as shown in
Fig.~\ref{fig:R2_bread_UENonUnifDis}, where examples of the possible
positions of $K_{2}$ within $R_{2}$ are shown as dots. Note that
the considered $f_{Z_{2}}\left(z\right)$ is just an example of the
non-uniformly distributed UEs in $R_{2}$, which reflects a reasonable
network planning that BSs have been deployed at the center spots of
UE clusters. Other forms of the UE distribution function will not
affect the procedure of our analysis, only the approximation error
values may change with the choice of the UE distribution function.
Since we have shown in Subsection~\ref{sub:B2_UEUnifDis} that Rayleigh
fading is the worst case for the proposed analysis, we adopt the assumption
of Rayleigh fading in this subsection.

From (\ref{eq:sum_epslong}), we can evaluate the quality of approximating
$I_{2}$ as a Gaussian RV by checking the corresponding $\varepsilon$.
Like Table~\ref{tab:results_bread_UEUnifDis_Rayleigh}, the results
of $\varepsilon$ for this network scenario are tabulated for various
values of $r$ in Table~\ref{tab:results_bread_UENonUnifDis_Rayleigh}.
From Table~\ref{tab:results_bread_UENonUnifDis_Rayleigh}, we can
see that the values of $\varepsilon$ are small, i.e., below 0.01,
which indicates that the approximation of $I_{2}$ as a Gaussian RV
should be tight%
, as can be confirmed from the actual error values in Table~\ref{tab:results_bread_UENonUnifDis_Rayleigh}.

\begin{table}[H]
\begin{centering}
{\small{}\protect\caption{\label{tab:results_bread_UENonUnifDis_Rayleigh}Approximation errors
of the proposed analysis ($B=2$, non-uniformly distributed UEs with
the distribution function of $f_{Z_{2}}\left(z\right)=\frac{W}{\rho},z\in R_{2}$,
Rayleigh fading).}
}
\par\end{centering}{\small \par}

{\small{}\vspace{-0.4cm}
}{\small \par}

\begin{centering}
{\small{}}%
\begin{tabular}{|l|l|l|l|l|l|l|l|l|}
\hline
{\small{}$r$} & {\small{}$\varepsilon$} & {\small{}$\varepsilon_{1}+\varepsilon_{2}$} & {\small{}$\varepsilon_{1}$} & {\small{}$\varepsilon'_{1}+\varepsilon'_{2}$} & {\small{}$\varepsilon'_{1}$} & {\small{}$\mu_{Q_{2}}$} & {\small{}$\sigma_{Q_{2}}^{2}$} & {\small{}Actual error}\tabularnewline
\hline
\hline
{\small{}$0.01$} & {\small{}$4.9\times10^{-3}$} & {\small{}$4.0\times10^{-4}$} & {\small{}$4\times10^{-6}$} & {\small{}$4.5\times10^{-3}$} & {\small{}$4\times10^{-6}$} & {\small{}-97.13} & {\small{}205.1} & {\small{}$1.8\times10^{-3}$}\tabularnewline
\hline
{\small{}$0.02$} & {\small{}$5.3\times10^{-3}$} & {\small{}$7.1\times10^{-4}$} & {\small{}$4\times10^{-6}$} & {\small{}$4.6\times10^{-3}$} & {\small{}$4\times10^{-6}$} & {\small{}-100.6} & {\small{}207.9} & {\small{}$2.2\times10^{-3}$}\tabularnewline
\hline
{\small{}$0.04$} & {\small{}$4.9\times10^{-3}$} & {\small{}$4.0\times10^{-4}$} & {\small{}$4\times10^{-6}$} & {\small{}$4.5\times10^{-3}$} & {\small{}$4\times10^{-6}$} & {\small{}-103.2} & {\small{}214.4} & {\small{}$1.7\times10^{-3}$}\tabularnewline
\hline
\end{tabular}
\par\end{centering}{\small \par}

\vspace{-0.5cm}
\end{table}

\subsection{The Scenario with Multiple Interfering Cells\label{sub:The-Scenario-mult-cell}}

In this subsection, we apply the proposed framework to a more complex
network with \emph{practical deployment of multiple cells} and provide
the approximation of the UL interference distribution.

\begin{figure}[H]
\noindent \begin{centering}
\vspace{-0.5cm}
\includegraphics[width=7cm]{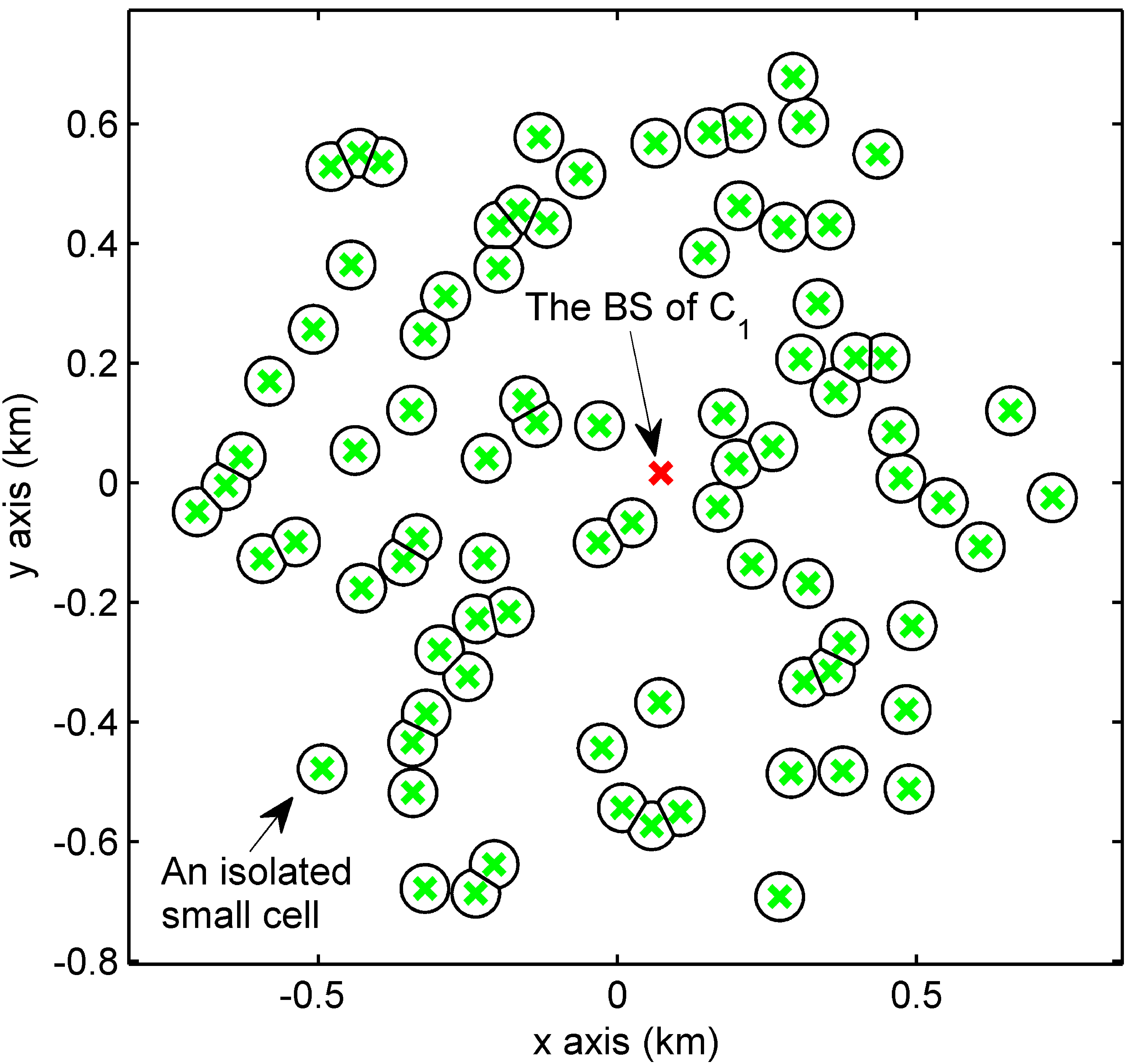}\renewcommand{\figurename}{Fig.}\protect\caption{\label{fig:3gpp_84cells}Illustration of a practical deployment of
multiple small cells ($r=0.04\,\textrm{km}$).}

\par\end{centering}

\vspace{-0.5cm}
\end{figure}

Here, we consider a 3GPP-compliant scenario~\cite{TR36.828}, as
shown in Fig.~\ref{fig:3gpp_84cells}, where the number of BSs $B$
is set to 84 and all small cell BSs are represented by x-markers.
Particularly, the BS of $C_{1}$ has been explicitly pointed out.
The reference coverage area for each small cell is a disk with a radius
of $r$~\cite{TR36.828}. As in previous subsections, the values
of $r$ (in km) are set to 0.01, 0.02 and 0.04, respectively. The
reference disk-shaped areas can be easily seen in Fig.~\ref{fig:3gpp_84cells}
from any isolated small cell. However, due to the irregular positions
of the cells%
, the actual coverage areas of the considered cells are of irregular
shapes. The irregularly shaped coverage areas are outlined by solid
lines in Fig.~\ref{fig:3gpp_84cells}. %

An important note is that the considered network scenario is different
from that adopted in~{[}2,3{]}, where coverage areas are defined
as Voronoi cells generated by the Poisson distributed BSs and the
\emph{entire} network area is covered by those Voronoi cells. In practice,
small cells are mainly used for capacity boosting in specific populated
areas, rather than the provision of an umbrella coverage for all UEs~\cite{Tutor_smallcell}.
In this light, the 3GPP standards recommend the hotspot scenario depicted
in Fig.~\ref{fig:3gpp_84cells} for UE distribution in the performance
evaluation of practical SCNs and we adopt such network scenario in
this subsection. %
Nevertheless, we should mention that the proposed microscopic analysis
of the UL interference distribution can still be applied on a particular
Voronoi tessellation. This is because Theorem~\ref{thm:Gauss+sth=00003DGauss}
and Corollary~\ref{cor:from_main_thm} in this paper do not rely
on particular shape and/or size of coverage areas.

In this subsection, we consider the same UE distribution function
as the one discussed in Subsection~\ref{sub:B2_UENonUnifDis}, i.e.,
$f_{Z_{b}}\left(z\right)=\frac{W}{\rho},z\in R_{b}$, where $\rho$
is the radial coordinate of $z$ in the polar coordinate system with
its origin placed at the position of the BS of $C_{b}$, and $W$
is a normalization constant to make $\int_{R_{b}}f_{Z_{b}}\left(z\right)dz=1$.
Besides, we assume Rayleigh fading in this subsection since Rayleigh
fading is the worst case for the proposed analysis as addressed in
Subsection~\ref{sub:B2_UEUnifDis}.

In the following, we investigate the considered network with the proposed
microscopic analysis of the UL interference distribution. First, for
each $R_{b},b\in\left\{ 2,\ldots,84\right\} $, we invoke (\ref{eq:epslong_1}),
(\ref{eq:epslong_2}) and (\ref{eq:epslong_prime}) to check the maximum
error $\varepsilon$ among $b\in\left\{ 2,\ldots,84\right\} $ using
(\ref{eq:sum_epslong}). If the maximum value of $\varepsilon$ is
reasonably small, e.g., less than 0.01, then we can approximate $I^{\textrm{mW}}$
in~(\ref{eq:rx_interf_UL}) as a power lognormal RV $\hat{I}^{\textrm{mW}}=10^{\frac{1}{{10}}Q}$,
where the PDF and the CDF of $Q$ are given by (\ref{eq:PDF_CDF_powerLN_Q})
with the parameters $\lambda$, $\mu_{Q}$ and $\sigma_{Q}^{2}$ obtained
from, e.g., Appendix~B.

The maximum values of the 83 $R_{b}$-specific $\varepsilon$'s for
various $r$ values are presented in Table~\ref{tab:results_84cells_hotspot_UENonUnifDis}.
From Table~\ref{tab:results_84cells_hotspot_UENonUnifDis}, we can
observe that, for all the investigated values of $r$, the maximum
values of $\varepsilon$ are below 0.01. %
Thus, each $I_{b}$ should be well approximated by a Gaussian RV $Q_{b}$.
Due to space limitation, we omit the detailed numerical investigation
on the Gaussian approximation for each $I_{b}$, which is very similar
to the discussion in Subsection~\ref{sub:B2_UENonUnifDis}. %
After obtaining the approximation for each $I_{b}$, we approximate
$I^{\textrm{mW}}$ in~(\ref{eq:rx_interf_UL}) as a power lognormal
RV $\hat{I}^{\textrm{mW}}=10^{\frac{1}{{10}}Q}$ using (\ref{eq:PDF_CDF_agg_Interf}).
The numerical results of $\lambda$, $\mu_{Q}$ and $\sigma_{Q}^{2}$
are provided in Table~\ref{tab:results_84cells_hotspot_UENonUnifDis}
for reference.

\begin{table}[H]
\begin{centering}
{\small{}\protect\caption{\label{tab:results_84cells_hotspot_UENonUnifDis}Approximation errors
of the proposed analysis ($B=84$, non-uniformly distributed UEs with
the distribution function of $f_{Z_{b}}\left(z\right)=\frac{W}{\rho},z\in R_{b}$,
Rayleigh fading).}
}
\par\end{centering}{\small \par}

{\small{}\vspace{-0.4cm}
}{\small \par}

\begin{centering}
{\small{}}%
\begin{tabular}{|l|l|l|l|l|l|l|l|l|}
\hline
{\small{}$r$} & {\small{}Max $\varepsilon$} & {\small{}Max $\varepsilon_{1}+\varepsilon_{2}$} & {\small{}Max $\varepsilon_{1}$} & {\small{}Max $\varepsilon'_{1}+\varepsilon'_{2}$} & {\small{}Max $\varepsilon'_{1}$} & {\small{}$\lambda$} & {\small{}$\mu_{Q}$} & {\small{}$\sigma_{Q}^{2}$}\tabularnewline
\hline
\hline
{\small{}$0.01$} & {\small{}$5.1\times10^{-3}$} & {\small{}$2.7\times10^{-4}$} & {\small{}$4\times10^{-6}$} & {\small{}$4.8\times10^{-3}$} & {\small{}$4\times10^{-6}$} & {\small{}48.9} & {\small{}-99.7} & {\small{}116.2}\tabularnewline
\hline
{\small{}$0.02$} & {\small{}$5.1\times10^{-3}$} & {\small{}$4.2\times10^{-4}$} & {\small{}$4\times10^{-6}$} & {\small{}$4.7\times10^{-3}$} & {\small{}$4\times10^{-6}$} & {\small{}48.0} & {\small{}-101.6} & {\small{}116.5}\tabularnewline
\hline
{\small{}$0.04$} & {\small{}$5.9\times10^{-3}$} & {\small{}$1.2\times10^{-3}$} & {\small{}$4\times10^{-6}$} & {\small{}$4.7\times10^{-3}$} & {\small{}$4\times10^{-6}$} & {\small{}47.5} & {\small{}-103.1} & {\small{}117.2}\tabularnewline
\hline
\end{tabular}
\par\end{centering}{\small \par}

\vspace{-0.5cm}
\end{table}

\begin{figure}[H]
\noindent \begin{centering}
\vspace{-0.5cm}
\includegraphics[width=8cm]{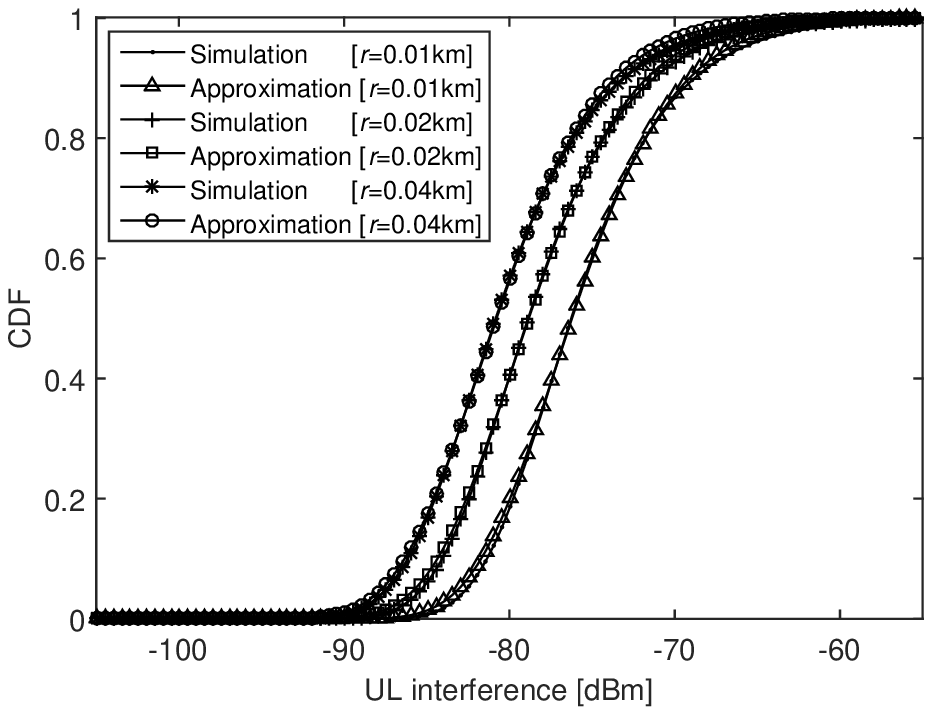}\renewcommand{\figurename}{Fig.}\protect\caption{\label{fig:UL_interf_approx_84cells_hotspot_UENonUnifDis_Rayleigh}The
simulation and the approximation of the CDF of $I^{\textrm{mW}}$
in dBm ($B=84$, non-uniformly distributed UEs with the distribution
function of $f_{Z_{b}}\left(z\right)=\frac{W}{\rho},z\in R_{b}$,
Rayleigh fading).}

\par\end{centering}

\vspace{-0.5cm}
\end{figure}

To further verify the accuracy of our analytical results on the UL
interference distribution, in Fig.~\ref{fig:UL_interf_approx_84cells_hotspot_UENonUnifDis_Rayleigh}
we plot the simulation results of the CDF of $I^{\textrm{mW}}$ in
dBm and the approximate analytical results according to (\ref{eq:PDF_CDF_agg_Interf}).
As can be seen from Fig.~\ref{fig:UL_interf_approx_84cells_hotspot_UENonUnifDis_Rayleigh},
the resulting power lognormal approximation of $I^{\textrm{mW}}$
is tight. However, note that the approximation shown in Fig.~\ref{fig:UL_interf_approx_84cells_hotspot_UENonUnifDis_Rayleigh}
is not as perfect as that exhibited in Fig.~\ref{fig:UL_interf_approx_irregR2_UEUnifDis_Rayleigh}.
According to the discussion in Section~\ref{sec:Analysis-of-UL-Interf},
the approximation errors associated with the first and the second
steps of approximation are captured in $\varepsilon$, which is very
small as can be confirmed from Table~\ref{tab:results_84cells_hotspot_UENonUnifDis}.
The noticeable small approximation errors in Fig.~\ref{fig:UL_interf_approx_84cells_hotspot_UENonUnifDis_Rayleigh}
are caused by the inaccuracy of approximating the sum of multiple
lognormal RVs as a single power lognormal RV%
. Note that in our previous work~\cite{UL_interference_approx_GC},
we use a lognormal RV to approximate the sum of multiple lognormal
RVs, which leads to larger errors compared with the results shown
in Fig.~\ref{fig:UL_interf_approx_84cells_hotspot_UENonUnifDis_Rayleigh}.
Finding an even better distribution to approximate the CDF of the
sum of multiple lognormal RVs will be our future work%
.

{} %

\subsection{Discussion on the Complexity of the Proposed Microscopic Analysis\label{sub:Further-Discussion}}

The computational complexity of the proposed approach is mainly attributable
to the numerical integration required to obtain the values of $\varepsilon$
for each small cell. In contrast, the simulation approach, e.g.,~\cite{UL_interf_sim1},~\cite{UL_interf_sim2},
as well as in this paper, involves a tremendously high complexity.
Specifically, in our simulations, in order to go through the randomness
of all the RVs discussed in Section~\ref{sec:Network-Model}, more
than \emph{one billion} of realizations of $I_{b}$ have been conducted
for the 83 interfering cells depicted in Fig.~\ref{fig:3gpp_84cells}.
This shows that the proposed microscopic analysis of network performance
is computationally efficient, which makes it a convenient tool to
study future 5G systems with general and dense small cell deployments.
Furthermore, our analytical studies yield better insight into the
performance of the system compared with simulations.

\section{Conclusion\label{sec:Conclusion}}

The lognormal approximation of the UL inter-cell interference in FDMA
SCNs is important because it allows tractable network performance
analysis%
. Compared with the existing works%
, in this work we have analytically derived an upper bound on the
error of such approximation, measured by the KS distance between the
actual CDF and the approximate CDF.

Our results are very general in the sense that we do not pose any
requirement on (i) the shape and/or size of cell coverage areas, (ii)
the uniformity of UE distribution, and (iii) the type of multi-path
fading. Based on our results, we have proposed a new approach to directly
and analytically investigate a complex network with \emph{practical
deployment of multiple BSs placed at irregular locations}, using the
approximation of the aggregate UL interference by a power lognormal
distribution. From our theoretical analysis and simulation results,
we can see that the proposed approach possesses the following merits:
\begin{enumerate}
\item It quantifies the approximation error measured by an upper-bound KS
distance using a closed-form function. And the tightness of the approximation
is validated by the numerical results.
\item It tolerates more practical assumptions than the existing works, e.g.,
irregular hot-spots, overlapped cells, etc. And it can cope with a
large number of small cells with a low computational complexity of
analysis, thus making it a convenient tool to study future 5G systems
with general and dense small cell deployments.
\end{enumerate}

As future work, we will further investigate the impact of the correlated
shadow fading, the three-dimensional (3D) antenna pattern and the
multi-antenna transmission on the proposed approximation of the UL
interference distribution.

\section*{Appendix A: Proof of Theorem~\ref{thm:Gauss+sth=00003DGauss}\label{sec:Appendix-A}}

For clarity, we first summarize our approach to prove Theorem~\ref{thm:Gauss+sth=00003DGauss}
as follows. Our idea is to perform a Fourier series expansion for
both the CDF of $\tilde{G}$ and that of the hypothetically approximate
Gaussian RV. The distance between those two CDFs will be quantified
by the upper-bound KS distance derived in closed-form expressions.

First, according to the definition of $\tilde{G}$, the CDF of $\tilde{G}$
can be formally represented by%

\noindent
\begin{eqnarray}
F_{\tilde{G}}\left(g\right)\hspace{-0.3cm} & = & \hspace{-0.3cm}\Pr\left[\tilde{G}\leq g\right]\nonumber \\
\hspace{-0.3cm} & = & \hspace{-0.3cm}\Pr\left[\tilde{L}+\tilde{S}\leq g\right]\nonumber \\
\hspace{-0.3cm} & = & \hspace{-0.3cm}\Pr\left[\tilde{S}\leq g-\tilde{L}\right]\nonumber \\
\hspace{-0.3cm} & \overset{\triangle}{=} & \hspace{-0.3cm}\int_{-\infty}^{+\infty}\Phi\left(\frac{g-l}{\sigma_{S}}\right)f_{\tilde{L}}\left(l\right)dl,\label{eq:CDF_Z_def}
\end{eqnarray}

\noindent where %
$\Phi\left(\cdot\right)$ is the CDF of the standard normal distribution.
According to~\cite{Book_math_tables}, $\Phi\left(\frac{g-l}{\sigma_{S}}\right)$
can be further written as $\Phi\left(\frac{g-l}{\sigma_{S}}\right)=1-\frac{1}{2}\textrm{erfc}\left(\frac{g-l}{\sqrt{2}\sigma_{S}}\right)$,
where $\textrm{erfc}\left(x\right)$ is the complementary error function
defined as $\textrm{erfc}\left(x\right)=\frac{2}{\sqrt{\pi}}\int_{x}^{\infty}\exp\left(-t^{2}\right)dt$
in~\cite{Book_Proakis}.

Second, due to the independence of $\tilde{L}$ and $\tilde{S}$,
the variance of the approximate Gaussian RV of $\left(\tilde{L}+\tilde{S}\right)$
should be $\mu_{\tilde{L}}+\mu_{\tilde{S}}=0$ and $\sigma_{L}^{2}+\sigma_{S}^{2}$,
respectively. As a result, the CDF of such approximate Gaussian RV
can be expressed as

\noindent
\begin{equation}
\Phi\left(\frac{g}{\sqrt{\sigma_{L}^{2}+\sigma_{S}^{2}}}\right)=1-\frac{1}{2}\textrm{erfc}\left(\frac{g}{\sqrt{2\left(\sigma_{L}^{2}+\sigma_{S}^{2}\right)}}\right).\label{eq:CDF_Z_Gauss}
\end{equation}

In the following, we will derive %
the closed-form expressions of the upper-bound KS distance between
the two CDFs presented in (\ref{eq:CDF_Z_def}) and (\ref{eq:CDF_Z_Gauss}),
respectively.

According to~\cite{erfc_Fourier_expansion}, $\textrm{erfc}\left(x\right)$
can be expanded to a Fourier series as

\noindent
\begin{equation}
\textrm{erfc}\left(x\right)=1-\frac{4}{\pi}\sum_{n=1,n\textrm{ odd}}^{N=2p-1}\frac{\exp\left(-n^{2}\omega^{2}\right)}{n}\sin\left(2n\omega x\right)+\delta\left(x\right),\label{eq:erfc_Fourier_expansion}
\end{equation}

\noindent where $\omega$ is the fundamental frequency of the Fourier
series, $N=2p-1$ is the series truncation point, and $\delta\left(x\right)$
is the residual error of the $p$-truncated Fourier series.

\noindent %

Based on (\ref{eq:erfc_Fourier_expansion}), $\Phi\left(\frac{g-l}{\sigma_{S}}\right)$
in (\ref{eq:CDF_Z_def}) can be expanded as%

\noindent
\begin{eqnarray}
\hspace{-0.3cm}\Phi\left(\frac{g-l}{\sigma_{S}}\right)\hspace{-0.3cm} & = & \hspace{-0.3cm}\frac{1}{2}+\frac{2}{\pi}\hspace{-0.2cm}\sum_{n=1,n\textrm{ odd}}^{2p-1}\hspace{-0.2cm}\frac{\exp\left(-n^{2}\omega^{2}\right)}{n}\sin\left(2n\omega\frac{g-l}{\sqrt{2}\sigma_{S}}\right)-\frac{1}{2}\delta\left(\frac{g-l}{\sqrt{2}\sigma_{S}}\right)\nonumber \\
\hspace{-0.3cm} & = & \hspace{-0.3cm}\frac{1}{2}+\frac{2}{\pi}\hspace{-0.2cm}\sum_{n=1,n\textrm{ odd}}^{2p-1}\hspace{-0.2cm}\frac{\exp\left(-n^{2}\omega^{2}\right)}{n}\textrm{imag}\left\{ \hspace{-0.1cm}\exp\hspace{-0.1cm}\left(\textrm{j}\sqrt{2}n\omega\frac{g-l}{\sigma_{S}}\right)\hspace{-0.1cm}\right\} -\frac{1}{2}\delta\left(\frac{g-l}{\sqrt{2}\sigma_{S}}\right),\label{eq:g(ql)_rewritten}
\end{eqnarray}

\noindent where $\textrm{imag}\left\{ \cdot\right\} $ extracts the
imaginary part of a complex value. %
Plugging (\ref{eq:g(ql)_rewritten}) into (\ref{eq:CDF_Z_def}), yields%

\noindent
\begin{eqnarray}
\hspace{-0.3cm}F_{\tilde{G}}\left(g\right)\hspace{-0.3cm} & = & \hspace{-0.3cm}\frac{1}{2}-\frac{1}{2}\int_{-\infty}^{+\infty}\delta\left(\frac{g-l}{\sqrt{2}\sigma_{S}}\right)f_{\tilde{L}}\left(l\right)dl\nonumber \\
\hspace{-0.3cm} &  & \hspace{-0.3cm}+\frac{2}{\pi}\textrm{imag}\hspace{-0.1cm}\left\{ \hspace{-0.1cm}\sum_{n=1,n\textrm{ odd}}^{2p-1}\hspace{-0.3cm}\exp\hspace{-0.1cm}\left(\textrm{j}\frac{\sqrt{2}n\omega g}{\sigma_{S}}\hspace{-0.1cm}\right)\hspace{-0.1cm}\frac{\exp\left(-n^{2}\omega^{2}\right)}{n}\hspace{-0.2cm}\int_{-\infty}^{+\infty}\hspace{-0.3cm}\exp\hspace{-0.1cm}\left(-\textrm{j}\frac{\sqrt{2}n\omega l}{\sigma_{S}}\hspace{-0.1cm}\right)\hspace{-0.1cm}f_{\tilde{L}}\left(l\right)dl\hspace{-0.1cm}\right\} ,\label{eq:immediate_results_1}
\end{eqnarray}

\noindent where the integral $\int_{-\infty}^{+\infty}\exp\left(-\textrm{j}\frac{\sqrt{2}n\omega l}{\sigma_{S}}\right)f_{\tilde{L}}\left(l\right)dl$
is actually the characteristic function~\cite{Book_math_tables}
of $f_{\tilde{L}}\left(l\right)$ evaluated at the point $t=-\frac{\sqrt{2}n\omega}{\sigma_{S}}$.
Such characteristic function, denoted by $\varphi_{\tilde{L}}\left(t\right)$,
is

\noindent
\begin{equation}
\varphi_{\tilde{L}}\left(t\right)\overset{\triangle}{=}\int_{-\infty}^{+\infty}\exp\left(\textrm{j}tl\right)f_{\tilde{L}}\left(l\right)dl.\label{eq:integration_Fourier_transform}
\end{equation}

Hence, from (\ref{eq:immediate_results_1}) and (\ref{eq:integration_Fourier_transform}),
(\ref{eq:CDF_Z_def}) can be re-formulated as

\noindent
\begin{eqnarray}
F_{\tilde{G}}\left(g\right)\hspace{-0.3cm} & = & \hspace{-0.3cm}\frac{1}{2}-\frac{1}{2}\int_{-\infty}^{+\infty}\delta\left(\frac{g-l}{\sqrt{2}\sigma_{S}}\right)f_{\tilde{L}}\left(l\right)dl\nonumber \\
\hspace{-0.3cm} &  & \hspace{-0.3cm}+\frac{2}{\pi}\textrm{imag}\left\{ \sum_{n=1,n\textrm{ odd}}^{2p-1}\exp\left(\textrm{j}\frac{\sqrt{2}n\omega g}{\sigma_{S}}\right)\frac{1}{n}\exp\left(-n^{2}\omega^{2}\right)\varphi_{\tilde{L}}\left(-\frac{\sqrt{2}n\omega}{\sigma_{S}}\right)\right\} \nonumber \\
\hspace{-0.3cm} & = & \hspace{-0.3cm}\frac{1}{2}-\frac{1}{2}\int_{-\infty}^{+\infty}\hspace{-0.1cm}\delta\left(\frac{g-l}{\sqrt{2}\sigma_{S}}\right)\hspace{-0.1cm}f_{\tilde{L}}\left(l\right)dl+\frac{2}{\pi}\textrm{imag}\left\{ \hspace{-0.1cm}\sum_{n=1,n\textrm{ odd}}^{2p-1}\hspace{-0.1cm}\upsilon_{n}\exp\left(\textrm{j}\frac{\sqrt{2}n\omega g}{\sigma_{S}}\right)\hspace{-0.1cm}\right\} ,\label{eq:CDF_Z_reform}
\end{eqnarray}

\noindent where $\upsilon_{n}$ is defined as

\noindent
\begin{equation}
\upsilon_{n}\overset{\triangle}{=}\frac{1}{n}\exp\left(-n^{2}\omega^{2}\right)\varphi_{\tilde{L}}\left(-\frac{\sqrt{2}n\omega}{\sigma_{S}}\right).\label{eq:vn}
\end{equation}
And the summation in the $\textrm{imag}\left\{ \cdot\right\} $ operator
of (\ref{eq:CDF_Z_reform}) can be deemed as the weighted sum of the
unit-amplitude complex values $\left\{ \exp\left(\textrm{j}\frac{\sqrt{2}n\omega g}{\sigma_{S}}\right)\right\} $,
$n\in\left\{ 1,3,\ldots,2p-1\right\} $. %
{}

Regarding $\upsilon_{n}$, by means of the series representation of
$\exp\left(\cdot\right)$, we can re-write it as

\noindent
\begin{eqnarray}
\upsilon_{n}\hspace{-0.3cm} & = & \hspace{-0.3cm}\frac{1}{n}\exp\left(-n^{2}\omega^{2}\right)\varphi_{\tilde{L}}\left(-\frac{\sqrt{2}n\omega}{\sigma_{S}}\right)\nonumber \\
\hspace{-0.3cm} & = & \hspace{-0.3cm}\frac{1}{n}\exp\left(-n^{2}\omega^{2}\right)\int_{-\infty}^{+\infty}\exp\left(-\textrm{j}\frac{\sqrt{2}n\omega l}{\sigma_{S}}\right)f_{\tilde{L}}\left(l\right)dl\nonumber \\
\hspace{-0.3cm} & = & \hspace{-0.3cm}\frac{1}{n}\exp\left(-n^{2}\omega^{2}\right)\int_{-\infty}^{+\infty}\left[1-\textrm{j}\frac{\sqrt{2}n\omega l}{\sigma_{S}}-\frac{1}{2!}\left(\frac{\sqrt{2}n\omega l}{\sigma_{S}}\right)^{2}\right.\nonumber \\
\hspace{-0.3cm} &  & \hspace{-0.3cm}\left.+\textrm{j}\frac{1}{3!}\left(\frac{\sqrt{2}n\omega l}{\sigma_{S}}\right)^{3}+\frac{1}{4!}\left(\frac{\sqrt{2}n\omega l}{\sigma_{S}}\right)^{4}+\cdots\right]f_{\tilde{L}}\left(l\right)dl\nonumber \\
\hspace{-0.3cm} & = & \hspace{-0.3cm}\frac{1}{n}\exp\left(-n^{2}\omega^{2}\right)\left[1-0-\frac{2n^{2}\omega^{2}\sigma_{L}^{2}}{2\sigma_{S}^{2}}+\textrm{j}\frac{\rho_{L}^{\left(3\right)}}{6}\left(\frac{\sqrt{2}n\omega}{\sigma_{S}}\right)^{3}+\frac{\rho_{L}^{\left(4\right)}}{24}\left(\frac{\sqrt{2}n\omega}{\sigma_{S}}\right)^{4}+\cdots\right]\nonumber \\
\hspace{-0.3cm} & \overset{\left(a\right)}{=} & \hspace{-0.3cm}\frac{1}{n}\exp\left(-n^{2}\omega^{2}\right)\exp\left(-\frac{n^{2}\omega^{2}\sigma_{L}^{2}}{\sigma_{S}^{2}}\right)+\kappa_{n}\nonumber \\
\hspace{-0.3cm} & = & \hspace{-0.3cm}\frac{1}{n}\exp\left(-n^{2}\omega^{2}\left(\frac{\sigma_{L}^{2}+\sigma_{S}^{2}}{\sigma_{S}^{2}}\right)\right)+\kappa_{n}\nonumber \\
 & = & \hspace{-0.3cm}\hat{\upsilon}_{n}+\kappa_{n},\label{eq:nu_n_approx}
\end{eqnarray}

\noindent where $\hat{\upsilon}_{n}$ is defined as

\noindent
\begin{equation}
\hat{\upsilon}_{n}\overset{\triangle}{=}\frac{1}{n}\exp\left(-n^{2}\omega^{2}\left(\frac{\sigma_{L}^{2}+\sigma_{S}^{2}}{\sigma_{S}^{2}}\right)\right).\label{eq:vn_hat}
\end{equation}
Besides, in (\ref{eq:nu_n_approx}), $\kappa_{n}$ measures the difference
between $\upsilon_{n}$ and $\hat{\upsilon}_{n}$, defined as $\kappa_{n}=\upsilon_{n}-\hat{\upsilon}_{n}$.
The motivation of introducing $\hat{\upsilon}_{n}$ into (\ref{eq:nu_n_approx})
is because $\left(1-\frac{n^{2}\omega^{2}\sigma_{L}^{2}}{\sigma_{S}^{2}}\right)$
can be approximated as $\exp\left(-\frac{n^{2}\omega^{2}\sigma_{L}^{2}}{\sigma_{S}^{2}}\right)$
when $\frac{n^{2}\omega^{2}\sigma_{L}^{2}}{\sigma_{S}^{2}}$ is small
in the step (a) of (\ref{eq:nu_n_approx}). %
Note that no approximation is assumed in (\ref{eq:nu_n_approx}) because
$\kappa_{n}$ fully captures the difference between $\upsilon_{n}$
and $\hat{\upsilon}_{n}$.

Plugging (\ref{eq:nu_n_approx}) and (\ref{eq:vn_hat}) into (\ref{eq:CDF_Z_reform}),
we can get%

\noindent
\begin{eqnarray}
F_{\tilde{G}}\left(g\right)\hspace{-0.3cm} & = & \hspace{-0.3cm}\frac{1}{2}-\frac{1}{2}\int_{-\infty}^{+\infty}\delta\left(\frac{g-l}{\sqrt{2}\sigma_{S}}\right)f_{\tilde{L}}\left(l\right)dl+\frac{2}{\pi}\textrm{imag}\left\{ \sum_{n=1,n\textrm{ odd}}^{2p-1}\kappa_{n}\exp\left(\textrm{j}\frac{\sqrt{2}n\omega g}{\sigma_{S}}\right)\right\} \nonumber \\
\hspace{-0.3cm} &  & \hspace{-0.3cm}+\frac{2}{\pi}\sum_{n=1,n\textrm{ odd}}^{2p-1}\frac{1}{n}\exp\left(-n^{2}\omega^{2}\left(\frac{\sigma_{L}^{2}+\sigma_{S}^{2}}{\sigma_{S}^{2}}\right)\right)\sin\left(\frac{\sqrt{2}n\omega g}{\sigma_{S}}\right).\label{eq:CDF_Z_reform_approx}
\end{eqnarray}

Performing a variable change of $\omega=\bar{\omega}\sqrt{\frac{\sigma_{S}^{2}}{\sigma_{L}^{2}+\sigma_{S}^{2}}}$
in (\ref{eq:CDF_Z_reform_approx}), we can obtain%

\noindent
\begin{eqnarray}
F_{\tilde{G}}\left(g\right)\hspace{-0.3cm} & = & \hspace{-0.3cm}\frac{1}{2}-\frac{1}{2}\int_{-\infty}^{+\infty}\delta\left(\frac{g-l}{\sqrt{2}\sigma_{S}}\right)f_{\tilde{L}}\left(l\right)dl+\frac{2}{\pi}\textrm{imag}\left\{ \sum_{n=1,n\textrm{ odd}}^{2p-1}\kappa_{n}\exp\left(\frac{\textrm{j}\sqrt{2}n\bar{\omega}g}{\sqrt{\sigma_{L}^{2}+\sigma_{S}^{2}}}\right)\right\} \nonumber \\
\hspace{-0.3cm} &  & \hspace{-0.3cm}+\frac{2}{\pi}\sum_{n=1,n\textrm{ odd}}^{2p-1}\frac{1}{n}\exp\left(-n^{2}\bar{\omega}^{2}\right)\sin\left(\frac{\sqrt{2}n\bar{\omega}g}{\sqrt{\sigma_{L}^{2}+\sigma_{S}^{2}}}\right).\label{eq:CDF_Z_reform_approx_variable_replace}
\end{eqnarray}

\noindent %

On the other hand, from (\ref{eq:erfc_Fourier_expansion}), (\ref{eq:CDF_Z_Gauss})
can also be expanded to a Fourier series using $\bar{\omega}$ as%

\noindent
\begin{eqnarray}
\Phi\left(\frac{g}{\sqrt{\sigma_{L}^{2}+\sigma_{S}^{2}}}\right)\hspace{-0.3cm} & = & \hspace{-0.3cm}\frac{1}{2}-\frac{1}{2}\delta\left(\frac{g}{\sqrt{2\left(\sigma_{L}^{2}+\sigma_{S}^{2}\right)}}\right)\nonumber \\
\hspace{-0.3cm} &  & \hspace{-0.3cm}+\frac{2}{\pi}\sum_{n=1,n\textrm{ odd}}^{2p-1}\frac{1}{n}\exp\left(-n^{2}\bar{\omega}^{2}\right)\sin\left(\frac{\sqrt{2}n\bar{\omega}g}{\sqrt{\sigma_{L}^{2}+\sigma_{S}^{2}}}\right),\label{eq:CDF_Z_reform_approx_Fourier_expansion}
\end{eqnarray}

\noindent where $\delta\left(\cdot\right)$ is the residual error
function incurred from the $p$-truncated Fourier series expansion
and it has been defined in (\ref{eq:erfc_Fourier_expansion}).

From (\ref{eq:CDF_Z_reform_approx_variable_replace}) and (\ref{eq:CDF_Z_reform_approx_Fourier_expansion}),
we can bound the distance between $F_{\tilde{G}}\left(g\right)$ and
$\Phi\left(\frac{g}{\sqrt{\sigma_{L}^{2}+\sigma_{S}^{2}}}\right)$
as%
{}

\noindent
\begin{eqnarray}
\left|F_{\tilde{G}}\left(g\right)-\Phi\left(\frac{g}{\sqrt{\sigma_{L}^{2}+\sigma_{S}^{2}}}\right)\right|\hspace{-0.3cm} & = & \hspace{-0.3cm}\left|-\frac{1}{2}\int_{-\infty}^{+\infty}\delta\left(\frac{g-l}{\sqrt{2}\sigma_{S}}\right)f_{\tilde{L}}\left(l\right)dl+\frac{1}{2}\delta\left(\frac{g}{\sqrt{2\left(\sigma_{L}^{2}+\sigma_{S}^{2}\right)}}\right)\right.\nonumber \\
\hspace{-0.3cm} &  & \hspace{-0.3cm}\left.+\frac{2}{\pi}\textrm{imag}\left\{ \sum_{n=1,n\textrm{ odd}}^{2p-1}\kappa_{n}\exp\left(\frac{\textrm{j}\sqrt{2}n\bar{\omega}g}{\sqrt{\sigma_{L}^{2}+\sigma_{S}^{2}}}\right)\right\} \right|\nonumber \\
 & \leq & \frac{1}{2}\int_{-\infty}^{+\infty}\left|\delta\left(\frac{g-l}{\sqrt{2}\sigma_{S}}\right)\right|f_{\tilde{L}}\left(l\right)dl+\frac{1}{2}\left|\delta\left(\frac{g}{\sqrt{2\left(\sigma_{L}^{2}+\sigma_{S}^{2}\right)}}\right)\right|\nonumber \\
 &  & +\frac{2}{\pi}\sum_{n=1,n\textrm{ odd}}^{2p-1}\left|\kappa_{n}\right|.\label{eq:KS_dis_pre_eq}
\end{eqnarray}

The first two terms in the right-hand side of (\ref{eq:KS_dis_pre_eq})
are caused by the residual errors from the $p$-truncated Fourier
series expansion. From~\cite{erfc_Fourier_expansion_error_bound},
$\delta\left(x\right)$ in (\ref{eq:erfc_Fourier_expansion}) can
be strictly bounded by

\noindent
\begin{equation}
\left|\delta\left(x\right)\right|<\frac{2}{\sqrt{\pi}\omega}\textrm{erfc}\left(\left(2p+1\right)\omega\right)+\textrm{erfc}\left(\frac{\pi}{2\omega}-\left|x\right|\right).\label{eq:error_bound_erfc_Fourier_expansion}
\end{equation}

\noindent Note that such bound is dependent on the value of $x$,
which makes it difficult to handle the first term in the right-hand
side of (\ref{eq:KS_dis_pre_eq}), because the integral in that term
is computed from $-\infty$ to $+\infty$. Thus, in the following,
we will discuss the upper-bound of $\left|\delta\left(x\right)\right|$
respectively for the case of $\left|x\right|\leq k$ and for the case
of $\left|x\right|>k$, where $k$ is a positive scalar%
.

Considering the monotonic increase of $\textrm{erfc}\left(\frac{\pi}{2\omega}-\left|x\right|\right)$
with respect to $\left|x\right|$, we can further bound $\left|\delta\left(x\right)\right|$
for a range of interest of $\left|x\right|<k$ as

\noindent
\begin{equation}
\left|\delta\left(x\right)\right|<\frac{2}{\sqrt{\pi}\omega}\textrm{erfc}\left(\left(2p+1\right)\omega\right)+\textrm{erfc}\left(\frac{\pi}{2\omega}-k\right)\overset{\triangle}{=}\delta_{0}\left(\omega,p,k\right),\quad\left(\left|x\right|<k\right).\label{eq:error_bound_erfc_Fourier_expansion_with_k}
\end{equation}

\noindent For the range of $\left|x\right|>k$, since $\textrm{erfc}\left(x\right)\leq2$,
$\left|\delta\left(x\right)\right|$ can be bounded by

\noindent
\begin{equation}
\left|\delta\left(x\right)\right|<\frac{2}{\sqrt{\pi}\omega}\textrm{erfc}\left(\left(2p+1\right)\omega\right)+2\overset{\triangle}{=}\delta_{1}\left(\omega,p\right),\quad\left(\left|x\right|\geq k\right).\label{eq:error_bound_erfc_Fourier_expansion_larger_than_k}
\end{equation}

In order to derive an upper bound that is independent of $g$ for
the right-hand side of (\ref{eq:KS_dis_pre_eq}), in the following
we consider two cases for $g$, i.e., $\left|g\right|<k_{1}\sqrt{\sigma_{L}^{2}+\sigma_{S}^{2}}$
and $\left|g\right|\geq k_{1}\sqrt{\sigma_{L}^{2}+\sigma_{S}^{2}}$,
where $k_{1}$ is an arbitrary positive scalar.

When $\left|g\right|<k_{1}\sqrt{\sigma_{L}^{2}+\sigma_{S}^{2}}$,
we can bound the first term in the right-hand side of (\ref{eq:KS_dis_pre_eq})
by

\noindent
\begin{eqnarray}
\frac{1}{2}\int_{-\infty}^{+\infty}\left|\delta\left(\frac{g-l}{\sqrt{2}\sigma_{S}}\right)\right|f_{\tilde{L}}\left(l\right)dl\hspace{-0.3cm} & < & \hspace{-0.3cm}\frac{1}{2}\delta_{1}\left(\omega,p\right)\hspace{-0.3cm}\underset{\left|l\right|\geq k_{2}\sqrt{\sigma_{L}^{2}+\sigma_{S}^{2}}}{\int}\hspace{-0.3cm}f_{\tilde{L}}\left(l\right)dl\nonumber \\
\hspace{-0.3cm} &  & \hspace{-0.3cm}+\frac{1}{2}\delta_{0}\left(\omega,p,\frac{\left(k_{1}+k_{2}\right)\sqrt{\sigma_{L}^{2}+\sigma_{S}^{2}}}{\sqrt{2}\sigma_{S}}\right)\hspace{-0.3cm}\underset{\left|l\right|<k_{2}\sqrt{\sigma_{L}^{2}+\sigma_{S}^{2}}}{\int}\hspace{-0.3cm}\hspace{-0.3cm}f_{\tilde{L}}\left(l\right)dl\nonumber \\
\hspace{-0.3cm} & < & \hspace{-0.3cm}\frac{1}{2}\delta_{1}\left(\omega,p\right)\frac{1}{k_{2}^{2}}+\frac{1}{2}\delta_{0}\left(\omega,p,\frac{\left(k_{1}+k_{2}\right)\sqrt{\sigma_{L}^{2}+\sigma_{S}^{2}}}{\sqrt{2}\sigma_{S}}\right),\label{eq:Fourier_error_part1}
\end{eqnarray}

\noindent where $k_{2}$ is another arbitrary positive scalar introduced
to facilitate the bounding of the integral from $-\infty$ to $+\infty$
with regard to $l$. The last step of (\ref{eq:Fourier_error_part1})
is valid because (i)

\noindent
\begin{equation}
\underset{\left|l\right|\geq k_{2}\sqrt{\sigma_{L}^{2}+\sigma_{S}^{2}}}{\int}\hspace{-0.3cm}f_{\tilde{L}}\left(l\right)dl=\textrm{Pr}\left[\left|\tilde{L}\right|\geq k_{2}\sqrt{\sigma_{L}^{2}+\sigma_{S}^{2}}\right]\leq\textrm{Pr}\left[\left|\tilde{L}\right|\geq k_{2}\sqrt{\sigma_{L}^{2}}\right]\leq\frac{1}{k_{2}^{2}},\label{eq:interm_result1}
\end{equation}

\noindent which comes from Chebyshev's inequality~\cite{Chebyshev_inequality},
and (ii) $\underset{\left|l\right|<k_{2}\sqrt{\sigma_{L}^{2}+\sigma_{S}^{2}}}{\int}\hspace{-0.3cm}\hspace{-0.3cm}f_{\tilde{L}}\left(l\right)dl<\hspace{-0.3cm}\underset{l\in\mathbb{R}}{\int}f_{\tilde{L}}\left(l\right)dl=1$.

Besides, the second term in the right-hand side of (\ref{eq:KS_dis_pre_eq})
can be bounded by

\noindent
\begin{equation}
\frac{1}{2}\left|\delta\left(\frac{g}{\sqrt{2\left(\sigma_{L}^{2}+\sigma_{S}^{2}\right)}}\right)\right|<\frac{1}{2}\delta_{0}\left(\bar{\omega},p,\frac{k_{1}}{\sqrt{2}}\right)=\frac{1}{2}\delta_{0}\left(\frac{\omega\sqrt{\sigma_{L}^{2}+\sigma_{S}^{2}}}{\sigma_{S}},p,\frac{k_{1}}{\sqrt{2}}\right),\label{eq:Fourier_error_part2}
\end{equation}
Plugging (\ref{eq:Fourier_error_part1}) and (\ref{eq:Fourier_error_part2})
into (\ref{eq:KS_dis_pre_eq}), and considering the definition of
$\varepsilon_{1}$ and $\varepsilon_{2}$ in (\ref{eq:epslong_1})
and (\ref{eq:epslong_2}), respectively, we can get

\noindent
\begin{equation}
\left|F_{\tilde{G}}\left(g\right)-\Phi\left(\frac{g}{\sqrt{\sigma_{L}^{2}+\sigma_{S}^{2}}}\right)\right|<\varepsilon_{1}+\varepsilon_{2},\quad\left(\left|g\right|<k_{1}\sqrt{\sigma_{L}^{2}+\sigma_{S}^{2}}\right).\label{eq:KS_dis_ineq_case1}
\end{equation}

When $\left|g\right|\geq k_{1}\sqrt{\sigma_{L}^{2}+\sigma_{S}^{2}}$,
we invoke Chebyshev's inequality~\cite{Chebyshev_inequality} to
obtain

\noindent
\begin{equation}
\textrm{Pr}\left[\left|\tilde{G}\right|\geq k_{1}\sqrt{\sigma_{L}^{2}+\sigma_{S}^{2}}\right]=\textrm{Pr}\left[\left(\tilde{G}\geq k_{1}\sqrt{\sigma_{L}^{2}+\sigma_{S}^{2}}\right)\bigcup\left(\tilde{G}\leq-k_{1}\sqrt{\sigma_{L}^{2}+\sigma_{S}^{2}}\right)\right]\leq\frac{1}{k_{1}^{2}}.\label{eq:Chebyshev_ineq}
\end{equation}

\noindent Since $\textrm{Pr}\left[\tilde{G}\geq k_{1}\sqrt{\sigma_{L}^{2}+\sigma_{S}^{2}}\right]=1-F_{\tilde{G}}\left(g\right)$
and $\left\{ \tilde{G}\leq-k_{1}\sqrt{\sigma_{L}^{2}+\sigma_{S}^{2}}\right\} \neq\textrm{Ø}$,
we have

\noindent
\begin{equation}
1-F_{\tilde{G}}\left(g\right)<\frac{1}{k_{1}^{2}},\quad\left(\left|g\right|\geq k_{1}\sqrt{\sigma_{L}^{2}+\sigma_{S}^{2}}\right).\label{eq:Chebyshev_ineq2}
\end{equation}

\noindent Therefore, we can bound $\left|F_{\tilde{G}}\left(g\right)-\Phi\left(\frac{g}{\sqrt{\sigma_{L}^{2}+\sigma_{S}^{2}}}\right)\right|$
$\left(\left|g\right|\geq k_{1}\sqrt{\sigma_{L}^{2}+\sigma_{S}^{2}}\right)$
by

\noindent
\begin{eqnarray}
\left|F_{\tilde{G}}\left(g\right)-\Phi\left(\frac{g}{\sqrt{\sigma_{L}^{2}+\sigma_{S}^{2}}}\right)\right|\hspace{-0.3cm} & = & \hspace{-0.3cm}\left|1-F_{\tilde{G}}\left(g\right)-\left[1-\Phi\left(\frac{g}{\sqrt{\sigma_{L}^{2}+\sigma_{S}^{2}}}\right)\right]\right|\nonumber \\
\hspace{-0.3cm} & \leq & \hspace{-0.3cm}1-F_{\tilde{G}}\left(g\right)+\left|1-\Phi\left(\frac{g}{\sqrt{\sigma_{L}^{2}+\sigma_{S}^{2}}}\right)\right|\nonumber \\
\hspace{-0.3cm} & < & \hspace{-0.3cm}\frac{1}{k_{1}^{2}}+\frac{1}{2}\textrm{erfc}\left(k_{1}\right),\quad\left(\left|g\right|\geq k_{1}\sqrt{\sigma_{L}^{2}+\sigma_{S}^{2}}\right),\label{eq:KS_dis_ineq_case2}
\end{eqnarray}

\noindent where (\ref{eq:Chebyshev_ineq2}) and $0\leq1-\Phi\left(\frac{g}{\sqrt{\sigma_{L}^{2}+\sigma_{S}^{2}}}\right)\leq\frac{1}{2}\textrm{erfc}\left(k_{1}\right)$
for $\left|g\right|\geq k_{1}\sqrt{\sigma_{L}^{2}+\sigma_{S}^{2}}$,
have been plugged into the second last step of (\ref{eq:KS_dis_ineq_case2}).

Combining (\ref{eq:KS_dis_ineq_case1}) and (\ref{eq:KS_dis_ineq_case2}),
and considering the definition of $\varepsilon_{3}$ in (\ref{eq:epslong_3}),
we can obtain (\ref{eq:ineq_thm_Gauss}), which concludes our proof.

\section*{Appendix B: An Example to Obtain $\lambda$, $\mu_{Q}$ and $\sigma_{Q}$\label{sec:Appendix-B}}

According to~\cite{power_LN_approx_GC}, with regard to the RV $Q$,
we have

\noindent
\begin{equation}
\left\{ \begin{array}{l}
\underset{q\rightarrow+\infty}{\lim}\frac{\partial\Phi^{-1}\left(F_{Q}\left(q\right)\right)}{\partial q}=\frac{1}{\sigma_{Q}}\\
\underset{q\rightarrow-\infty}{\lim}\frac{\partial\Phi^{-1}\left(F_{Q}\left(q\right)\right)}{\partial q}=\frac{\sqrt{\lambda}}{\sigma_{Q}}
\end{array}\right..\label{eq:eqs_PLN_tail_express1}
\end{equation}

Besides, according to \cite{Approx_sumLN} and \cite{para_deci_power_LN_approx},
we can get

\noindent
\begin{equation}
\left\{ \begin{array}{l}
\underset{q\rightarrow+\infty}{\lim}\frac{\partial\Phi^{-1}\left(F_{Q}\left(q\right)\right)}{\partial q}=\frac{1}{\sigma_{X}}\\
\underset{q\rightarrow-\infty}{\lim}\frac{\partial\Phi^{-1}\left(F_{Q}\left(q\right)\right)}{\partial q}=\sqrt{\sum_{b=2}^{B}\sigma_{Q_{b}}^{-2}}
\end{array}\right.,\label{eq:eqs_PLN_tail_express2}
\end{equation}
where $\sigma_{X}$ is obtained by solving the following equation
set~\cite{Approx_sumLN}

\noindent
\begin{equation}
\left\{ \begin{array}{l}
{{\hat{\Psi}}_{X}}\left({s_{1}}\right)=\prod\limits _{b=2}^{B}{{{\hat{\Psi}}_{{Q_{b}}}}\left({s_{1}}\right)}\overset{\Delta}{=}C_{1}\\
{{\hat{\Psi}}_{X}}\left({s_{2}}\right)=\prod\limits _{b=2}^{B}{{{\hat{\Psi}}_{{Q_{b}}}}\left({s_{2}}\right)}\overset{\Delta}{=}C_{2}
\end{array}\right.,\label{eq:eqs_para_for_Gsum}
\end{equation}

\noindent where ${\hat{\Psi}}_{X}\left(s\right)$ is the approximate
moment generating function (MGF) evaluated at $s$ for a lognormal
RV defined as $10^{\frac{1}{{10}}X}$. Such approximate MGF is formulated
as

\noindent
\begin{equation}
{\hat{\Psi}}_{X}\left(s\right)=\hspace{-0.1cm}\sum\limits _{m=1}^{M_{0}}{\frac{{w_{m}}}{{\sqrt{\pi}}}\exp\left(\hspace{-0.1cm}{-s\exp\hspace{-0.1cm}\left(\hspace{-0.1cm}{\frac{{\sqrt{2\sigma_{X}^{2}}{a_{m}}+{\mu_{X}}}}{\zeta}}\right)}\hspace{-0.1cm}\right)},\label{eq:approx_MGF}
\end{equation}

\noindent where $\zeta=\frac{10}{\ln10}$, $M_{0}$ is the order of
the Gauss-Hermite numerical integration, the weights $w_{m}$ and
abscissas $a_{m}$ for $M_{0}$ up to 20 are tabulated in Table 25.10
in~\cite{GH_num_integration}. Usually, $M_{0}$ is set to be larger
than 8~\cite{Approx_sumLN}. Similarly, in (\ref{eq:eqs_para_for_Gsum}),
${\hat{\Psi}}_{Q_{b}}\left(s\right)$ is computed by replacing $\mu_{X}$
and $\sigma_{X}$ respectively with $\mu_{Q_{b}}$ and $\sigma_{Q_{b}}$
in (\ref{eq:approx_MGF}). In~(\ref{eq:eqs_para_for_Gsum}), $s_{1}$
and $s_{2}$ are two design parameters for generating two equations
that can determine the appropriate values of $\mu_{X}$ and $\sigma_{X}^{2}$.
For example, we can choose $s_{1}=0.001$, $s_{2}=0.005$ and $M_{0}=12$
as recommended in~\cite{Approx_sumLN}. The solution of (\ref{eq:eqs_para_for_Gsum})
can be readily found by standard mathematical software programs such
as MATLAB. %
Besides, using $\mu_{X}$ obtained by solving (\ref{eq:eqs_para_for_Gsum}),
we can match the mean of $Q$ with $\mu_{X}$ to construct a third
equation to determine the three parameters, i.e., $\lambda$, $\mu_{Q}$
and $\sigma_{Q}^{2}$, in the power lognormal distribution of $Q$.

To sum up, based on (\ref{eq:eqs_PLN_tail_express1}), (\ref{eq:eqs_PLN_tail_express2})
and matching the mean of $Q$ with $\mu_{X}$, we have the following
equation set to determine the values of $\lambda$, $\mu_{Q}$ and
$\sigma_{Q}$,

\noindent
\begin{equation}
\left\{ \begin{array}{ll}
\frac{1}{\sigma_{Q}}=\frac{1}{\sigma_{X}} & \quad(a)\\
\frac{\sqrt{\lambda}}{\sigma_{Q}}=\sqrt{\sum_{b=2}^{B}\sigma_{Q_{b}}^{-2}} & \quad(b)\\
\int_{-\infty}^{+\infty}qf_{Q}\left(q\right)dq=\mu_{X} & \quad(c)
\end{array}\right..\label{eq:eqs_para_for_Q}
\end{equation}

\noindent Note that equations (a) and (b) of (\ref{eq:eqs_para_for_Q})
are easy to solve and they can deliver $\lambda$ and $\sigma_{Q}$,
while equations (c) of (\ref{eq:eqs_para_for_Q}) can be efficiently
solved by the standard bisection process~\cite{Bisection} to determine
$\mu_{Q}$.

\end{document}